\documentclass[aps,12pt,axodraw,nofootinbib,superscriptaddress,aps]{revtex4}
\pdfoutput=1
\usepackage{epsfig}
\usepackage{amsmath}
\usepackage{bm}
\usepackage{times}
\usepackage{graphicx}
\usepackage{epstopdf}
\usepackage{amsfonts}
\usepackage{bm}
\usepackage{epsfig}
\usepackage{graphics}
\usepackage{xspace}
\usepackage[usenames]{color}

\def\nn{\nonumber}
\def\bea{\begin{eqnarray}}
\def\eea{\end{eqnarray}}
\def\ba{\begin{eqnarray}}
\def\ea{\end{eqnarray}}
\def\be{\begin{equation}}
\def\ee{\end{equation}}

\def\beq{\begin{equation}}
\def\eeq{\end{equation}}

\def\nn{\nonumber}

\begin{document}

\title{\Large First Glimpses at Higgs' face}
\author{J.R. Espinosa}
\affiliation{ICREA at IFAE, Universitat Aut{\`o}noma de Barcelona, 08193 Bellaterra, Barcelona, Spain}

\author{C. Grojean}
\affiliation{Theory Division, Physics Department, CERN, CH-1211 Geneva 23, Switzerland}

\author{M. M\"uhlleitner}
\affiliation{Institute for Theoretical Physics, Karlsruhe Institute of Technology, D-76128 Karlsruhe, Germany}

\author{M. Trott}
\affiliation{Theory Division, Physics Department, CERN, CH-1211 Geneva 23, Switzerland}
%
%

\date{\today}
\begin{abstract}
The $8 \, {\rm TeV}$ LHC Higgs search data just released indicates the 
existence of a scalar resonance with mass $ \sim 125 \, {\rm GeV}$.
We examine the implications of the data reported by ATLAS, CMS and the Tevatron collaborations on understanding the properties of this scalar by performing joint fits on its couplings to other Standard Model (SM) particles. We discuss and characterize to what degree this resonance has the properties of the SM Higgs, and consider what implications can be extracted for New Physics in a (mostly) model-independent fashion. We find that, if the Higgs couplings to fermions and weak vector bosons are allowed to differ from their standard values, the SM is $\sim 2\sigma$ from the best fit point to the current data. Fitting to a possible invisible decay branching ratio,
we find ${\rm BR}_{inv} \simeq 0.05\pm 0.32\ (95\%\  {\rm C.L.})$. We also discuss and develop some ways of using the data in order to bound or rule out models which modify significantly the properties of this scalar resonance, and apply these techniques to the  current global dataset.
\end{abstract}
\maketitle
\section{Introduction}
Particle physics entered a new era with  the announcement of the discovery of a new boson \cite{Wedtalk} based on excess events in several Higgs search channels using $7 + 8$ $\rm TeV$ LHC data collected in 2011-2012.
In light of this discovery, it has become obvious that the question of central importance to address now is -- what are the properties of the scalar field responsible for the observed excesses?
The answer to this question determines if this field corresponds to the Standard Model (SM) Higgs, with the specific SM mechanism of elegantly breaking electroweak (EW) symmetry, or whether a more complicated mechanism is involved in EW symmetry breaking. We study this question in detail in this paper, characterizing to what degree a SM Higgs is consistent with the current global dataset and presenting several results on the properties of the scalar field. Besides updating and expanding our past results \cite{Espinosa:2012ir}, we also present new analyses that emphasize the power of the growing dataset to bound and rule out alternative models or to give hints of New Physics (NP).  

It is worth emphasizing that it is very important to specify (and justify) a coherent theoretical framework in which to study the emerging evidence for 
the scalar field. However, without knowing the ultraviolet (UV) origin of this field, we do not know what effective field theory (EFT), or complete model, should be used to fit the data. 
We emphasize that at this time, the existing experimental evidence is not sufficiently strong to directly assume that the scalar resonance is the SM Higgs boson, ascribing any deviations in the measured properties of the scalar field directly to the effects of NP interactions expressed through higher dimensional operators. Although this is certainly one possible interpretation of the data (and we will examine the implications of Higgs data for NP in this framework), we emphasize that, in general, one should not assume what one wishes to prove.

Nevertheless, in formulating a theoretical framework for 
this study, a wealth of other experimental results that are also sensitive to the properties of scalar fields at the weak scale can be distilled into simple physical guiding principles. These are, namely, approximate
Minimal Flavour Violation (MFV) \cite{Chivukula:1987py,Hall:1990ac,D'Ambrosio:2002ex,Buras:2003jf,Cirigliano:2005ck}; respecting the soft Higgs theorems of Refs.~\cite{Shifman:1979eb,Vainshtein:1980ea} (i.e. the scalar couples to the SM fields in proportion to their masses); and an effective breaking of custodial symmetry, $\rm SU(2)_c$,  \cite{Susskind:1978ms,Weinberg:1979bn,Sikivie:1980hm} approximately as in the SM.
Directly incorporating these principles in the formulation of the effective Lagrangian allows us to restrict our attention to a few simple cases.
In order to establish experimentally the properties of the scalar resonance in a model-independent way, one can utilize the effective field theory of the chiral EW Lagrangian coupled to a scalar field 
that was emphasized in Refs.~\cite{Espinosa:2012ir,Azatov:2012bz} to study recent Higgs signal-strength data.\footnote{For other model-independent approaches to the determination of the Higgs couplings, see Refs.~\cite{Carmi:2012yp,Giardino:2012ww,Ellis:2012rx,Lafaye:2009vr,Englert:2011aa,Klute:2012pu,
concha,Low:2012rj,Carmi:2012zd,Azatov:2012ga}.} Depending on the assumptions of the UV origin of such a Lagrangian, one is lead to various sets of free parameters to fit the data when studying the consistency of the SM Higgs hypothesis with the current dataset.
We discuss and utilize this framework extensively in this paper to examine the properties of the scalar field emerging from the data.

We also emphasize that with the discovery of a new scalar resonance, one can also use the signal strength properties of the scalar field to bound and rule out models that
provide $\it too \, few$ signal events as well as models that provide too many signal events. Further, one can also exclude allowed parameter space due to the degree of tension within the dataset that depends on the properties of the scalar field.
As the dataset evolves, these techniques become complementary to direct $\chi^2$ fits on the signal strength dataset.
These bounds can be quantified by excluding parameter space in the allowed couplings of the scalar using a Gaussian probability density function approach. We develop and apply
such an approach in this paper.

The outline of this paper is as follows. In Section \ref{frameworks} we discuss the EFT framework we employ, and the implicit assumptions about the UV origin of the scalar field that are adopted when fitting the
data with various free parameters. In Section \ref{data} we review and discuss the manner in which we treat the scalar signal strength data and electroweak precision data (EWPD), while in Section \ref{results} we present results, based on our fit method,
of the status of the SM Higgs hypothesis. In Section \ref{BSM} we discuss some of the implications of Higgs signal strength parameters for beyond the SM (BSM) physics, expressed through model-independent free parameters.
In Section \ref{tension} we discuss novel and complementary methods to identify the allowed parameter space, and in Section \ref{concl} we conclude.
\section{Theoretical Framework}\label{frameworks}

An effective chiral EW Lagrangian with a nonlinear realization of the $\rm SU(2)_L \times U(1)_Y$ symmetry gives a minimal description of the (non-scalar) degrees of freedom of the SM  consistent with the assumptions of SM-like $\rm SU(2)_c$  violation and  MFV. The Goldstone bosons eaten by the $\rm
W^\pm, Z$ are denoted by $\pi^a$ (where
$a = 1,2,3$), and are grouped as 
\bea
\Sigma(x) = e^{i \sigma_a \, \pi^a/v} \; ,
\eea
with $\sigma_a$ the Pauli matrices and $v = 246 \, {\rm GeV}$.  
In this approach, the EW scale $v$,  which sets the mass of fermions and gauge bosons is introduced directly into the Lagrangian. The $\Sigma(x)$ field transforms
linearly under $\rm SU(2)_L \times SU(2)_R$ as $\Sigma(x) \rightarrow
L \, \Sigma(x) \, R^\dagger$ where $L,R$ indicate the transformation
on the left and right under $\rm SU(2)_L$ and $\rm SU(2)_R$, respectively, while $\rm SU(2)_c$ is the diagonal subgroup of
$\rm SU(2)_L \times SU(2)_R$. 

Adding a scalar field $h$ to this theory is trivial. One chooses $h$ to transform as a singlet under $\rm SU(2)_c$ and a derivative expansion of such a theory
is given by \cite{Giudice:2007fh,Contino:2010mh,Grober:2010yv}
\bea
\mathcal{L}_{eff} &=& \frac{1}{2} (\partial_\mu h)^2 - V(h) + \frac{v^2}{4} {\rm Tr} (D_\mu \Sigma^\dagger \, D^\mu \Sigma) \left[1 + 2 \, a \, \frac{h}{v} + b \, \frac{h^2}{v^2}  +  b_3 \, \frac{h^3}{v^3} + \cdots \right], \nn \\
&\,& - \frac{v}{\sqrt{2}} \, (\bar{u}_L^i \bar{d}_L^i) \, \Sigma \, \left[1 + c_j \, \frac{h}{v} +  c_2 \, \frac{h^2}{v^2}  + \cdots \right] 
\left(
\begin{array}{c} 
y_{ij}^u \, u_R^j \\ 
y_{ij}^d \, d_R^j 
\end{array} \right)  +  h.c. \cdots, 
\label{Leff}
\eea
with
\bea
V(h) &=& \frac{1}{2} \, m_h^2 \, h^2 + \frac{d_3}{6} \, \left(\frac{3
    \, m_h^2}{v} \right) \, h^3 + \frac{d_4}{24} \left(\frac{3
    m_h^2}{v^2} \right) h^4 + \cdots \; . 
\label{eq:efflag}
\eea
Although we use the notation $h$, we do not assume in principle that this scalar field is the Higgs,
or that the scale $v$ is somehow associated with the vacuum expectation value of this field -- as this is what we seek to establish from the data. As is well known, the $a$ and $c_j$ parameters control the couplings of $h$ to gauge bosons and fermions, respectively, and therefore, play a crucial role in the phenomenology of single $h$ production. Note that in previous fits, and in the majority of this work, the assumption $c_j \, y^{u,d}_{ij} \equiv c \, y^{u,d}_{ij}$ is used
and no distinction is made between the rescaling of the $h$-coupling to the $u^i$ and $d^i$ quarks.
We will relax this assumption later on.  This Lagrangian
is common in the study of composite models and its relevance has been emphasized recently in Refs.~\cite{Giudice:2007fh,Contino:2010mh,Grober:2010yv}. It is also appropriate to study a pseudo-Goldstone boson (PGB) emerging out of an approximately conformal sector \cite{Bando:1986bg,Goldberger:2007zk}, or as the low-energy EFT arising in many scenarios where the Higgs is a composite PGB that emerges from the breaking of a larger chiral symmetry group \cite{Kaplan:1983fs, Kaplan:1983sm,Agashe:2004rs,Barbieri:2007bh,Chang:2012gn,Chang:2012tb}. We emphasize that this EFT setup can be matched to many UV frameworks and, being quite general, we do not confine ourselves to any particular UV scenario.\footnote{The symmetry assumptions we adopt by  directly interpreting the data are minimal. In more involved scenarios, these assumptions can in principle be relaxed (see {\it e.g.} Ref.~\cite{Farina:2012ea}
for a study with this aim that relaxes $\rm SU(2)_c$ constraints). Note, however, that  relaxing the assumption $c_j \, y^{u,d}_{ij} \equiv c \, y^{u,d}_{ij} $ significantly, can lead to conflict with
precision constraints sensitive to $\rm SU(2)_c$ and flavour violation.}

We approach the data in this way to be as model-independent as possible. However, even specifying the free parameters that one will use to fit the data introduces
implicit model dependence.  One can nevertheless broadly characterize certain parameter choices. As this is an EFT, $\mathcal{L}_{eff}$ is non-renormalizable. Here (and throughout this paper) we take the cut-off scale to be $\Lambda \sim 4 \, \pi v /\sqrt{|1 - a^2|}$.  Since we are concerned with the
phenomenology of single scalar production, the higher-order derivative operators are suppressed by powers of $\mathcal{O}(m_h^2/\Lambda^2)$.
As such, we are justified in neglecting such sub-leading effects in this paper. 
Non-derivative higher-dimensional operators will also exist in general. When the $h$ field is not assumed to have a UV origin such as the SM Higgs, and is simply considered to be
a singlet field [that need not necessarily transform under the nonlinearly realized $\rm SU(2)_L \times U(1)$ symmetry], the leading operators in the expansion in inverse powers of $\Lambda$
appear at dimension five and are given by 
%
\bea
\mathcal{L}^5_{HD} &=& - \frac{c_g \, g_3^2}{2 \, \Lambda} \,h \, G^A_{\mu\, \nu} G^{A \, \mu \, \nu} - \frac{c_W \, g_2^2}{2 \, \Lambda} \, h \, W^a_{\mu\, \nu} W^{a \, \mu \, \nu} 
 - \frac{c_B \, g_1^2}{2 \, \Lambda} \, h \, B_{\mu\, \nu} B^{\mu \, \nu} \;.
\eea
Here $g_1,g_2,g_3$ are the weak hypercharge, $\rm {SU}(2)_L$  and $\rm{SU}(3)_c$ gauge couplings, respectively, and the different tensor fields are the corresponding field strengths with their associated Wilson coefficients $c_i$.
The scale $\Lambda$ corresponds to the mass scale of the lightest new state that is integrated out, which we assume is proximate to $\Lambda$. 
We will neglect operators originating from CP-violating sources due to the lack of any clear evidence of beyond the SM CP violation in lower energy precision tests.
Note that the operators in $\mathcal{L}^5_{HD}$ can be further suppressed compared to the effects of $a,c$ on (single) scalar production when the scalar field has specific UV origins. This is the case for example when $h$ is a PGB,
see Ref.~\cite{Giudice:2007fh} for a detailed discussion.

When one assumes that $h$ is embedded into an $\rm SU(2)_L$ doublet - $H$ - as in the SM, the operators in $\mathcal{L}^5_{HD}$  first appear at dimension six,
and the coefficients are suppressed by an extra factor of $v/\Lambda$ when considering single scalar production. In this case, the dimension six 
operator basis is also extended by the operator
\bea
\delta \mathcal{L}^6_{HD} &=&  - \frac{c_{WB} \, g_1 \, g_2}{2 \, \Lambda^2} \, H^\dagger \, \tau^a \, H \, B_{\mu\, \nu} W^{a \, \mu \, \nu} . 
\eea
For phenomenological purposes it is convenient to rotate to a basis for the operators given by
\bea\label{phenolagrangian}
\mathcal{L}_{HD} &=& - \frac{c_g \, g_3^2}{2 \, \Lambda} \,h \, G^A_{\mu\, \nu} G^{A \, \mu \, \nu} - \frac{c_{\gamma} \, (2 \, \pi \,  \alpha)}{\Lambda} \, h \, F_{\mu\, \nu} F^{\mu \, \nu},
\eea
where $F_{\mu \, \nu}$ is the electromagnetic field strength tensor and $c_{\gamma} = c_W +c_B$ in the case of an $\rm SU(2)_L$ singlet field, and $c_{\gamma} = c_W + c_B - c_{WB}$ if $h$ is embedded into an
$\rm SU(2)_L$ doublet. 

In this manner, one can understand that the choices to retain the effects of higher dimensional operators (or not) in performing global fits introduces implicit UV dependence.
In introducing higher dimensional operators, one is also explicitly assuming the existence of new states charged under at least a subgroup of the SM group. Alternatively, if NP is uncharged under the SM group but couples to the
$h^2$ operator\footnote{New physics of this form is sometimes referred to as coupling to the SM through the Higgs portal, see Refs.~\cite{Schabinger:2005ei,Patt:2006fw,MarchRussell:2008yu,Espinosa:2007qk,Pospelov:2011yp} for some related discussion.}, then it can impact $h$-phenomenology by inducing an invisible $h$-width (when the scalar field takes a vacuum expectation value). This leads to the modification of the SM branching ratios for each decay into visible SM final states $f$  via
\bea
{\rm Br}(h \rightarrow f) \equiv (1 - {\rm Br}_{inv}) \times {\rm Br}_{SM}(h \rightarrow f).
\label{eq:brinv}
\eea
We will update our recent fit of  ${\rm Br}_{inv}$ to Higgs search data \cite{Espinosa:2012vu} in a later section, and use this fit as a diagnostic tool to test aspects of our fit procedure. 

In general, the coefficients $a,c,c_{\gamma}, c_g, {\rm Br}_{inv} \cdots $ are arbitrary parameters subject to experimental constraints.
The generic cases we consider are:

\begin{itemize}
\item Composite/Pseudo-Goldstone Higgs/Dilaton scalar theories.

In this case, $a,c$ are free parameters in general, although they can be fixed in particular UV completions. If the scalar field is a Pseudo-Goldstone boson, it is also appropriate to neglect higher dimensional operators.
	We will fit to subsets of the parameters $\{a,c,\rm BR_{inv},c_{\gamma}, c_g\}$ in what follows. As this is a more general framework than
	the SM Higgs, we will use this EFT in assessing to what degree the SM Higgs hypothesis is consistent with the data or if deviations into this parameter space can give a substantially better fit.

	\item The SM Higgs as a low-energy EFT.

When the low-energy EFT is just the SM, the field $h$ becomes part of a linear multiplet
\bea
U = \left(1 + \frac{h}{v} \right) \Sigma  \; ,
\eea
reducing Eq.~(\ref{Leff}) to the SM Higgs Lagrangian.
In this case $a = b = c = d_3 = d_4 =1$ and $b_3 = c_2 =0$, and 
the only effect of NP is through non-renormalizable higher-order operators. The naturalness problem of the SM Higgs mass operator, and recent experimental hints of deviations in the observed properties of the (assumed) Higgs, motivates moderately heavy NP and the introduction of BSM parameters $(c_{\gamma}, c_g,\rm BR_{inv})$ as free parameters. We study the constraints on these parameters in detail in this paper.
\end{itemize}

We will not attempt to relate the constraints obtained on the various parameters to any particular underlying model, other than the SM, in this paper. This choice is motivated by the current lack of other clear experimental evidence of BSM states to guide coherent model-building. The classes of models discussed above can be considered as motivating examples.

\section{Data Treatment}\label{data}
\subsection{Signal-Strength Data}\label{mudata}

In this section we describe our method for globally fitting to the parameters discussed above, and incorporating the recently released $8$ TeV data \cite{Wedtalk}, updated $7$ TeV results from ATLAS \cite{atlasupdate}, the released 7 TeV CMS data 
\cite{Chatrchyan:2012tx}, and the recently reported Tevatron Higgs search results \cite{tevatronupdate}.
This work builds on our previous fits \cite{Espinosa:2012ir,Espinosa:2012vu}.
We only summarize the main details of the fit procedure
and method here. Many subsidiary details of the fit procedure can be found in these reference works. 

We fit to the available Higgs signal strength data, 
\bea
\mu_i = \frac{[ \sum_j \, \sigma_{j \rightarrow h} \times {\rm Br}(h \rightarrow i)]_{observed}}{[ \sum_j \, \sigma_{j \rightarrow h} \times {\rm Br}(h \rightarrow i)]_{SM}}\ ,
\label{mui}
\eea 
for the production of a Higgs  that decays into the visible channels $i = 1 \cdots N_{ch}$, where $N_{ch}$ denotes the number of channels. 
The label $j$ in the cross section, $\sigma_{j \rightarrow h}$,  is due to the fact that some final states are defined to only be summed over a subset of Higgs production processes $j$.
The reported best fit value of a signal strength we denote by $\hat{\mu}_i$\footnote{In a simple counting experiment one has $\hat\mu_i=(n_{obs,i}-n_{backg,i})/n_{s,i}^{SM}$, in terms of the observed numbers of events ($n_{obs,i}$), the number of background events ($n_{backg,i}$) and the expected number of SM signal events ($n_{s,i}^{SM}$).}.

The global $\chi^2$ we construct  is defined via
\bea
\label{chi2}
\chi^2(\mu_i) =\sum_{i=1}^{N_{ch}}\frac{(\mu_i-\hat\mu_i)^2}{\sigma_i^2}\;.
\eea 
The covariance matrix has been taken to be diagonal with the square of the $1 \, \sigma$ theory and experimental errors added in quadrature 
for each observable, giving the error $\sigma_i$ in the equation above. Correlation
coefficients are neglected as they are not supplied by the experimental collaborations. For the experimental errors we 
use $\pm$ symmetric $1 \sigma$ errors on the reported $\hat{\mu}_i$. For theory predictions of the $\sigma_{j \rightarrow h}$ and related errors, 
we use the numbers given on the webpage of the LHC Higgs Cross Section Working Group \cite{Dittmaier:2011ti}.\footnote{These values have recently been updated for $7,8$ TeV and we use the updated numbers.
Also note that ${\rm BR}(s \, \bar{s})$ is set to zero on this page but we use the latest version of HDECAY \cite{Djouadi:1997yw} to add in ${\rm BR}(s \, \bar{s})$ to the quoted results. This has a negligible impact on the reported numbers through
the modification of the total width.} 
The minimum ($\chi^2_{min}$) is determined, and the $68.2 \% \, (1 \, \sigma), 95 \% \, (2 \, \sigma), 99 \% \, (3 \, \sigma)$ best fit regions are plotted as $\chi^2 = \chi^2_{min} + \Delta \chi^2$, with
the appropriate cumulative distribution function (CDF) defining $ \Delta \chi^2$. 

We assume, as in Ref.~\cite{Azatov:2012bz,Espinosa:2012ir}, that 
the signal strength $\mu_i$ in a given channel $i$ follows a Gaussian distribution with the probability density function (pdf) given by
\bea
\mathrm{pdf}_i(\mu_i,\hat{\mu}_i,\sigma_i) \approx e^{-(\mu_i - \hat{\mu}_i)^2/(2 \sigma_i^2)},
\eea
with one-sigma error $\sigma_i$, and best fit value $\hat{\mu}_i$. This is the case as long as the number of events is large, $\gtrsim \mathcal{O}(10)$ events, \cite{Azatov:2012bz}. We normalize these pdf's to 1 in the interval $(0,\infty)$.

In the framework of the SM, the predicted values of the $\mu_i$ are the same (equal to 1), and a universal signal strength modifier $\mu$ can be defined and applied to all channels.
By multiplying together the individual channel pdf's (or the pdf's of a single channel reported at two operating energies), we can also define a combined PDF for $\mu$. This can be done for each separate experiment or for a global combination of all experiments. The combined PDF is also Gaussian and has combined $\hat\mu_c$ and $\sigma_c$ values given approximately\footnote{This neglects (unsupplied) correlations and is therefore a rough approximation that should be taken with due caution. An estimate of the accuracy of  this procedure can be done by comparing quantities derived from such combinations vs. the experimental ones, which typically agree within 5-10 \%. Generalizing (\ref{csigma}) to include correlations is straightforward.
These formulas are also easy to generalize in models with non-universal theory-predicted $\mu_i$.}  by
\be\label{csigma}
\frac{1}{\sigma_c^2}  = \sum_{i}^{N_{ch}} \frac{1}{\sigma^2_i} \, , \quad \quad  \frac{\hat{\mu}_c}{\sigma_c^2} = \sum_i^{N_{ch}} \frac{\hat \mu_i }{\sigma_i^2} \ . 
\ee
We will use these relations to reconstruct the unreported $8 \, {\rm TeV}$ data from the reported $7$ and $7+8 \, {\rm TeV}$ data.\footnote{In version two of this paper we have incorporated a number of refinements in our treatment of the data, affecting the results that follow in the body of the paper. These refinements are now possible due to
further information being released by the experimental collaborations after version one of this paper. See the Appendix for further details on the dataset now used.}%

Armed with combined PDF(s), we can determine the 95\% C.L. exclusion upper limits on the signal strength parameter $\mu$ ($\mu<\mu_{upL}$) \cite{Azatov:2012bz,Espinosa:2012ir}.
(We shall explain momentarily how to introduce an overall signal strength parameter in models with non-universal theory-predicted $\mu_i$. 
 In the discussion that follows we are implicitly assuming that 
we are considering setting limits on combined signal strength parameters although we frame the discussion in terms of $\mu$.
A similar analysis can  be carried through channel-by-channel instead of on the combined channels.)
Such limits can be set on the combined signal strength parameter $\mu_c$ or for an individual channel's signal strength. 
With $\hat\mu_i$'s settling around unity with $\sigma_i$  errors getting smaller and smaller due to increasing integrated luminosities, we can already start to set also {\it lower} limits on $\mu$. 
The condition for such lower bounds, say at 95$\%$ C.L.,  to be meaningful is that the symmetric interval $\hat\mu\pm\delta_{95}\hat\mu$
containing 95\% of the integrated probability has $\hat\mu-\delta_{95}\hat\mu>0$, in which case $\mu_{dwL}\equiv \hat\mu-\delta_{95}\hat\mu$ corresponds to  the lower 95\% C.L. bound on $\mu$. In this same case, we will take $\hat\mu+\delta_{95}\hat\mu>0$ as the upper limit
$\mu_{upL}$.
The conditions that define these limit are therefore
\be
\int^{\hat\mu}_{\mu_{dwL}} {\rm PDF}(\mu) d\mu =
\frac{\text{erf}\left[\frac{\text{$\hat\mu $}-\text{$\mu_{dwL} $}}{\sqrt{2} \sigma }\right]}{1+\text{erf}
\left[\frac{\text{$\hat\mu $}}{\sqrt{2} \sigma }\right]}=
 0.95/2, 
\ee
and
\be
\int_{\hat\mu}^{\mu_{upL}} {\rm PDF}(\mu) d\mu =
\frac{\text{erf}\left[\frac{\text{$\mu_{upL} $}-\text{$\hat{\mu}$}}{\sqrt{2} \sigma }\right]}{1+\text{erf}
\left[\frac{\text{$\hat\mu $}}{\sqrt{2} \sigma }\right]}=
 0.95/2\ ,
\ee
where ${\rm erf}(z)$ is the error function.
However, when $\hat\mu-\delta_{95}\hat\mu<0$, we shift the 95\% C.L. interval to the asymmetric one $(0,\mu_{upL})$
and revert to the upper limit definition  \cite{Azatov:2012bz,Espinosa:2012ir}\footnote{The 95\% C.L. interval extends down to 0 when the ratio $\sigma/\hat\mu> 0.6$, in which case there is no lower limit on $\mu$ and the upper limit is given by Eq.~(\ref{usualuplim}). }
\be
\label{usualuplim}
\int_0^{\mu_{upL}} {\rm PDF}(\mu) d\mu = \frac{\text{erf}\left[\frac{\text{$\hat\mu $}}{\sqrt{2} \sigma }\right]-\text{erf}\left[\frac{\text{$\hat\mu $}-\text{$\mu_{upL} $}}{\sqrt{2} \sigma }\right]}{1+\text{erf}\left[\frac{\text{$\hat\mu $}}{\sqrt{2} \sigma }\right]}= 0.95\ .
\ee

Next consider models that depart from the SM, like ${\rm SM}(a,c)$ with couplings of the Higgs to fermions and gauge bosons modified by the $a,c$ factors as explained above. The predicted values of the $\mu_i$'s will deviate from 1 in a channel-dependent way. In order to keep the same expected value of the signal-strengths for all channels, it is convenient to normalize the 
$\mu_i$'s in Eq.~(\ref{mui}) not to the SM signal expectation $n_{s,i}^{SM}$ but to the expectation in the model considered $n_{s,i}^{{\rm SM}(a,c)}$. By doing this we can again use a universal signal-strength modifier $\mu$, with expected value $\mu=1$, corresponding to the model being tested. On the other hand, with this change in the $\mu_i$ definition, the observed $\hat\mu_i \pm \sigma_i$ values need to be rescaled by a factor $n_{s,i}^{SM}/n_{s,i}^{{\rm SM}(a,c)}$. The individual pdf's are modified accordingly and can be combined in the usual way.

To summarize, we can exclude (at 95 \% C.L.) a given scenario not only if it predicts too many signal events which are not seen ($\mu>\mu_{upL}$), but also if it predicts too few events, incompatible with the observed excesses associated with the reported discovery. The significance of such lower bounds will grow with more luminosity. Significances above 5$\sigma$ become possible (by definition) after discovery
(which excludes the background hypothesis $\mu=0$). We will plot both of these bounds mapping the allowed  $\mu_{upL}^i,\mu_{dwL}^i$ into the relevant parameter space through the dependence of $\mu_i$ on the free parameters in our fit results. Figure \ref{tenslim}, in Section \ref{tension}, shows examples of such lower limits.

\subsection{Electroweak Precision Data}

We incorporate EWPD  \cite{Holdom:1990tc,Peskin:1991sw,Altarelli:1990zd} by adding it directly to the $\chi^2$ measure in Eq.~(\ref{chi2}). When $a$ is considered a free parameter, the shifts of the oblique parameters $S$ and $T$ are given by \cite{Barbieri:2007bh}
\bea
\Delta S &\approx& \frac{-(1 - a^2)}{6 \, \pi} \, \log \left(\frac{m_h}{\Lambda}\right),  \quad \quad  \Delta T \approx \frac{3(1 - a^2)}{8 \, \pi \, \cos^2 \theta_W} \, \log \left(\frac{m_h}{\Lambda}\right).
\eea
The numerical coefficient is determined from the logarithmic large-$m_h$ dependence of $S,T$ given in Ref.~\cite{Peskin:1991sw}.\footnote{
Here we have introduced an Euclidean momentum cut-off scale $\Lambda$. The degree to which
$\Lambda$ properly captures the UV regularization of $\rm S$ and $\rm T$ is model-dependent. We assume that directly treating this cut-off scale as a proxy for a heavy mass scale integrated out is a good approximation, i.e. that further arbitrary parameters rescaling the cut-off scale terms need not be introduced.} 
As for EWPD, recent updates to the measurement of $m_W$ at the Tevatron \cite{Aaltonen:2012bp,Abazov:2012bv} have refined the world average \cite{TevatronElectroweakWorkingGroup:2012gb},
and have significantly reduced the quoted error. Incorporating\footnote{We thank J. Erler for kindly providing these EWPD results.}  these new measurements we use \cite{jens}
\bea
S = 0.00 \pm 0.10,  \quad \quad  T = 0.02  \pm 0.11,  \quad \quad U = 0.03  \pm 0.09 \; ,
\eea
while the matrix of correlation coefficients is given by
\bea
C = \left(
\begin{array}{ccc} 
1 & 0.89 & -0.55 \\ 
0.89 & 1 & -0.80 \\
 -0.55 & -0.80 & 1 \\
\end{array} \right).
\eea
Here we have assumed $m_h = 125 \, {\rm GeV}$, as corrections for shifting these results by a few GeV are negligible. There is a strong preference for $a \simeq 1$ in the global fit when EWPD is used, and the constraints on the scalar field can be directly associated with EWPD bounds. Note that the slight preference for $a >1$ in the best fit region when EWPD is taken into account is subject to uncertainties in cut-off scale effects. Although the shift in the best fit point
is of interest as a probe of possible new physics, we cannot clearly disentangle such a hint from cut-off scale effects.

\section{Fit Results}\label{results}
\subsection{Status of the Higgs hypothesis}

The excess of events of $\approx 5\sigma$ significance reported by ATLAS and CMS peaks, as a function of the scalar mass, at slightly different values:
$126.5 \, {\rm GeV}$  for ATLAS and $125 \, {\rm GeV}$ for CMS.
This difference can be attributed at this stage of the search to statistical fluctuations in the data. Monte-Carlo studies \cite{Murray} indicate that such effects can shift the observed maximum signal strength compared to the true signal strength maximum by $\sim 2-3 \, {\rm GeV}$. Due to this, a fit that combines data from different experiments at the same mass value might be biased and not necessarily better than using data at slightly different masses.
As much more detailed data is available to us at the mass peaks, in our global fits we will use all available $\mu_i$  taken at $m_h = 125$ \, ${\rm GeV}$ for the CMS and the Tevatron (which has a $\approx 3\sigma$ excess over a wider region of masses), and at $m_h = 126.5 \, {\rm GeV}$ for ATLAS (see Fig.~\ref{Fig.data} in the Appendix).\footnote{Recent updates from the experimental collaborations that have appeared since version one of this paper have released more information at $m_h = 125.5 \, {\rm GeV}$ for CMS \cite{cmslatest} and at $m_h = 126 \,{\rm GeV}$ for ATLAS \cite{atlaslatest}. However, as the amount of information relevant to the fits we will perform released to date is still greater at the mass scales $m_h = 125 \,  {\rm GeV}$ for CMS and $m_h = 126.5 \,{\rm GeV}$ for ATLAS we retain these mass choices in our fit.}

\begin{figure}[tb]
\includegraphics[width=0.45\textwidth]{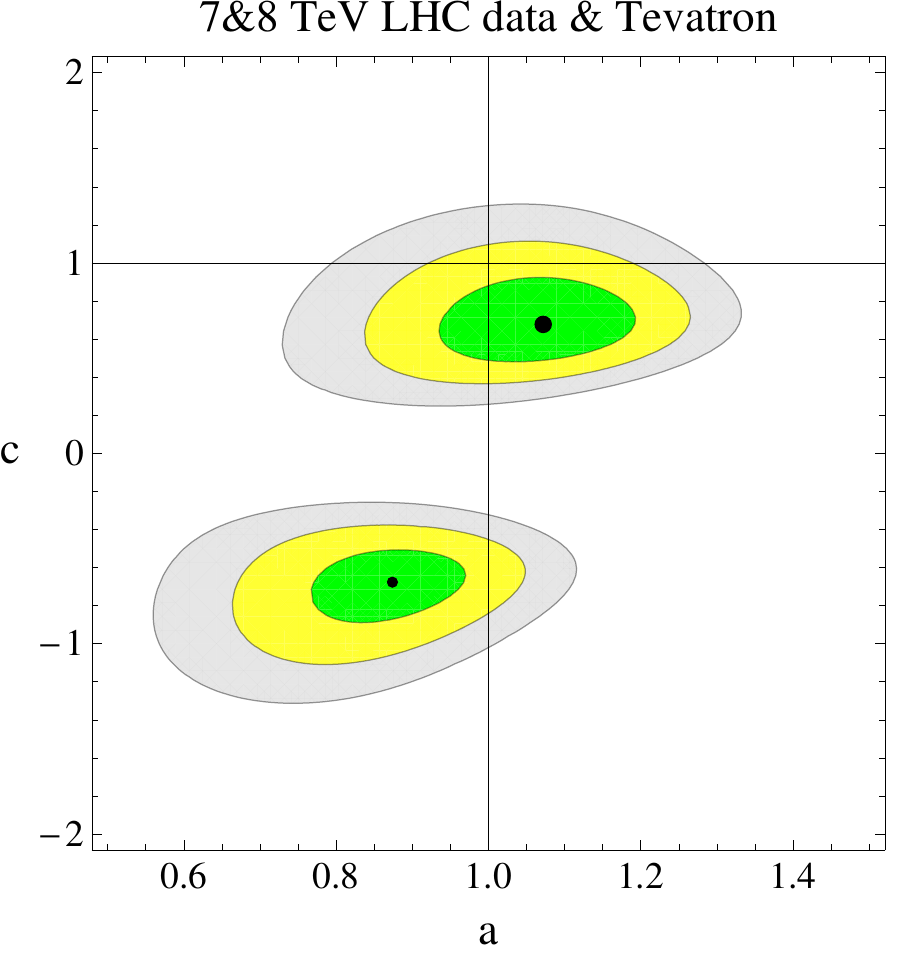}
\includegraphics[width=0.46\textwidth]{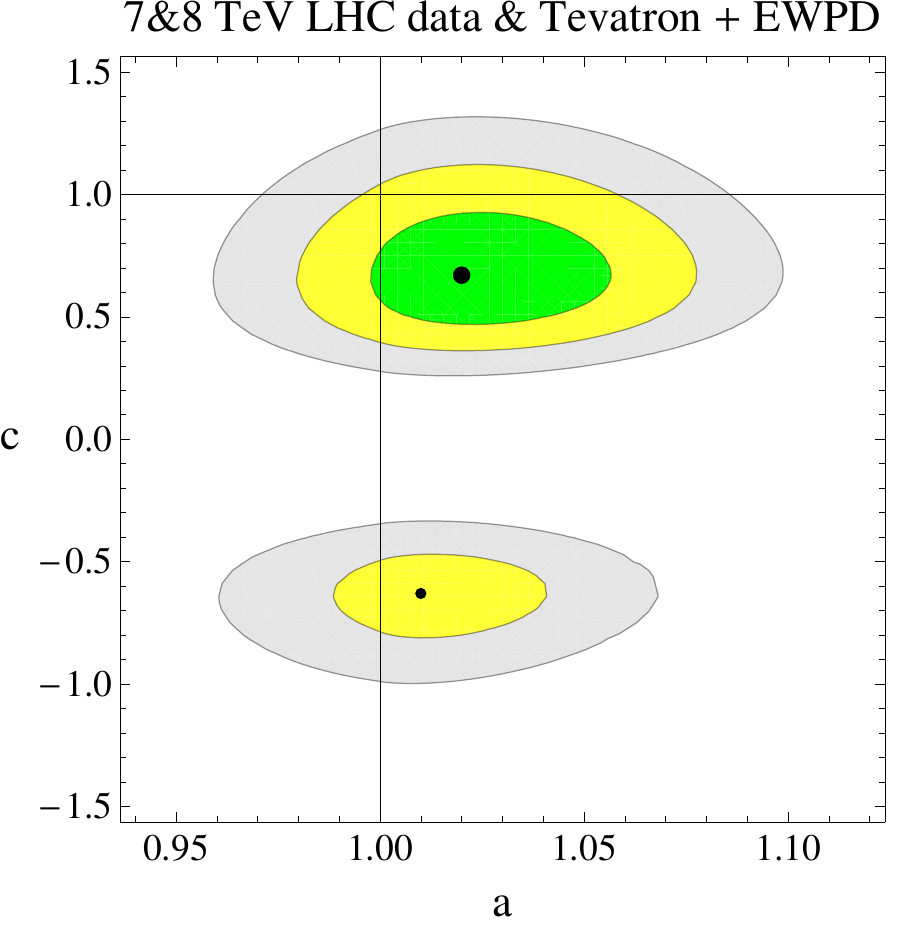}
\caption{Global fit results in the $(a,c)$ plane for all reported best fit values given by ATLAS and CMS, left (right) without EWPD (with EWPD). In both plots we take $m_h = 125 \, {\rm GeV}$ for the Tevatron and CMS7/8 and $m_h = 126.5 \, {\rm GeV}$ 
for ATLAS7/8. The green, yellow, gray regions corresponds to the allowed $1,2,3 \, \sigma$ spaces for a two parameter fit. The best fit point in each region is also labeled with a point. The thicker point indicates the one with the smaller $\chi^2_{min}$.}\label{bestfits}
\end{figure}

In order to assess the degree of consistency of the data with the SM hypothesis we first consider the effective Lagrangian given by $\mathcal{L}_{eff}$ and assume that higher dimensional operators are sufficiently suppressed so that $c_{\gamma}, c_g$ can be neglected.
We then perform a two-parameter $\chi^2(a,c)$ fit and examine the $\Delta \, \chi^2$
for the SM point $(a,c) = (1,1)$ compared with the best fit point. This defines a C.L. corresponding to the deviation of the SM hypothesis compared to the best fit point.
The result is shown in Figure \ref{bestfits} (left). We also show in Fig.~\ref{bestfits} (right) the best fit regions when EWPD is added to the global $\chi^2$ measure. Notice the dramatic reduction in the size of the best fit region along the $a$-parameter, which is forced to lie close to 1. These results visually summarize the current experimental status of establishing the Higgs hypothesis.

When EWPD is not used, the SM Higgs hypothesis of $(a,c)=(1,1)$ is $\lesssim 2 \, \sigma$ (C.L. of $0.88$) away from the best fit point, which sits at $(a,c) = (1.1,0.68)$. Note that here and in the following discussion we are choosing
to round the C.L.  and the best fit points. This is due to the preliminary nature of the $7+8\, {\rm TeV}$ data and is not limited directly to this accuracy due to the fit procedure we have adopted.
The C.L. of the SM hypothesis in the combined data
is consistent with our past results at $7$ $\rm TeV$ \cite{Espinosa:2012ir}.  With the updated data released and incorporated in our fit since version one of this paper
the parameter space that has the global minimum has changed from $c<0$ (with initial ICHEP data) to $c>0$ (with post-ICHEP updates).\footnote{The lack of data released for these subcategories to date at $m_h = 126 \, {\rm GeV}$ is the primary reason we retain the use of $m_h = 126.5 \, {\rm GeV}$ for ATLAS data.} The existence of the non negligible parameter space with $c<0$ is easy to understand. Due to the interference
term in the $h \rightarrow \gamma \, \gamma$ decay width which is $\propto -ac$, a negative $c$ allows a relatively larger excess in $\gamma \, \gamma$ events due to constructive interference between the top and $W$ boson loops.
When EWPD is used as in Figure \ref{bestfits} (right) we find that the SM is similarly residing at $\sim 2 \, \sigma$ (C.L. of $0.93$) away from the best fit point which is now
$(a,c) = (1.0,0.67)$ and the best fit region where $c>0$ now has a (significantly) lower global minimum. The minima are no longer as degenerate with the addition of the most recent ATLAS data,
$\Delta \chi^2(min_1,min_2) \sim 4$.

\begin{figure}[tb]
\includegraphics[width=0.32\textwidth]{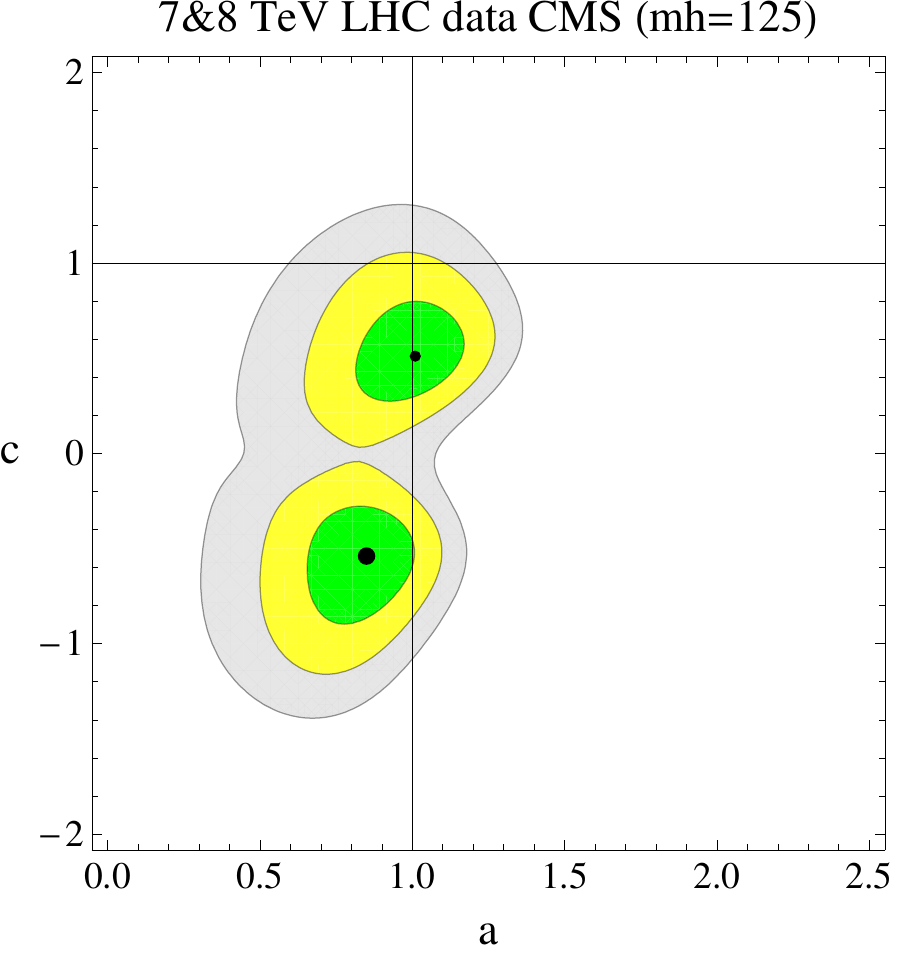}
\includegraphics[width=0.32\textwidth]{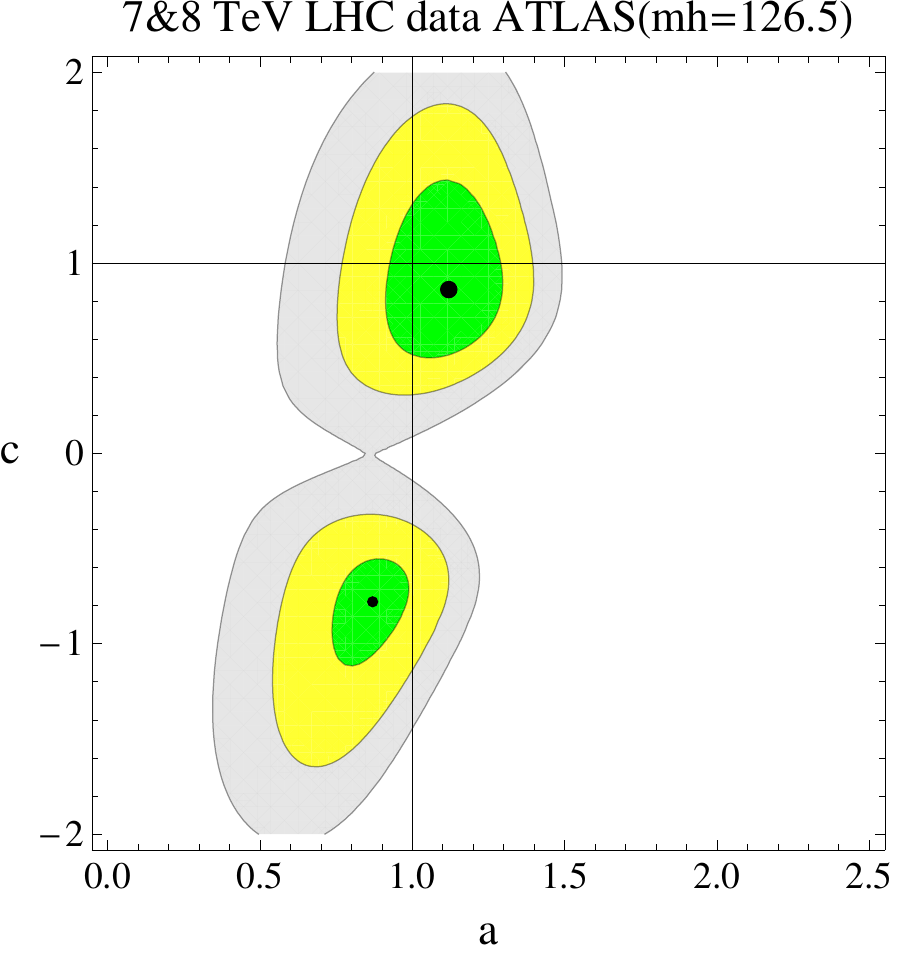}
\includegraphics[width=0.32\textwidth]{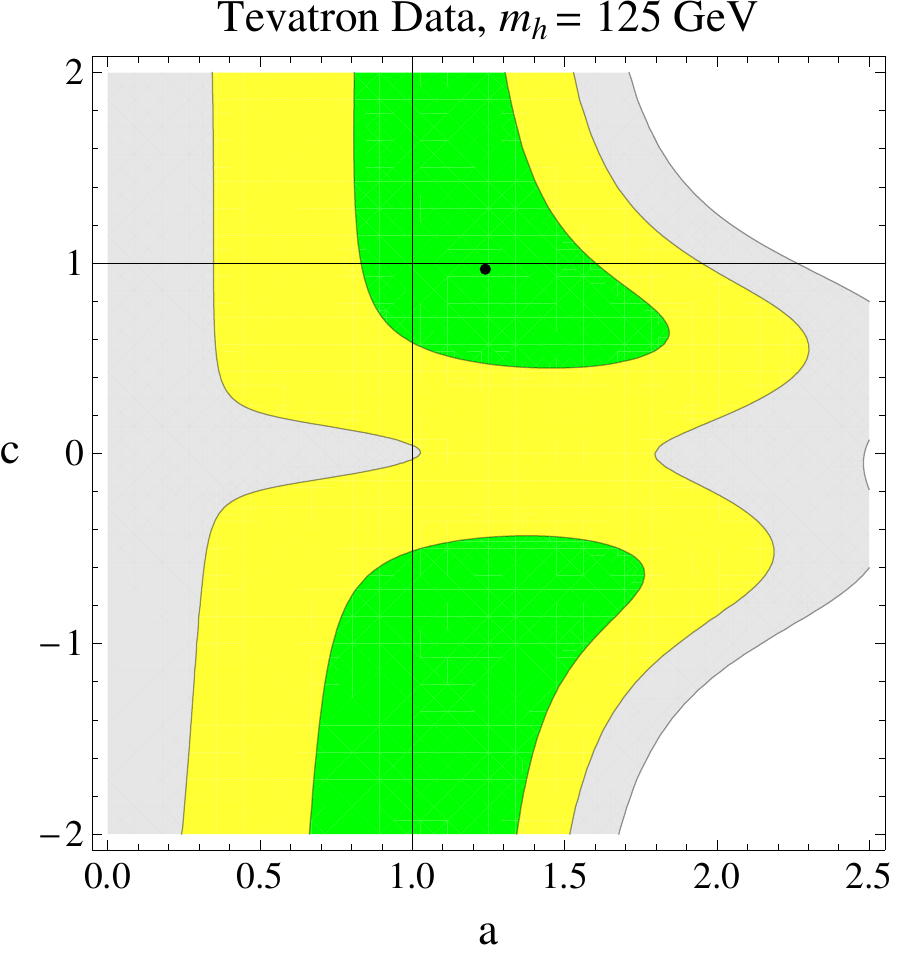}
\caption{Best fit regions (at 68\%, 95\% and 99.9\% C.L.) in the $(a,c)$ plane for a fit to all reported signal-strength values given by ATLAS ($m_h = 126.5 \, {\rm GeV}$), CMS ($m_h = 125 \, {\rm GeV}$) and the Tevatron ($m_h = 125 \, {\rm GeV}$) collaborations individually.
We plot the same best fit contours over the same domain of parameter space to allow a direct comparison amongst experimental results. The significant change in the ATLAS results from version one of this paper is due to the use of the ATLAS diphoton data broken into subcategories.}
\label{individualexp}
\end{figure}

In view of the different masses of the signal-strength peaks in the various experiments (which can be due to the statistical effects mentioned above) and of the subtleties we have neglected in properly combining the results of these different experiments, it is also of interest to perform the fit in the $(a,c)$ space for each experiment individually. We show these results in Figure \ref{individualexp}.
The CMS experiment has the SM point residing about $\sim 2 \sigma$ from the best fit point, with the C.L. of the SM case compared to the best fit point at $93 \%$. For
ATLAS, the SM point is now at a C.L. of  $41 \%$, within the $\sim 1 \sigma$ region. The Tevatron results have the SM point within the $1\sigma$ region with a C.L. of the SM case (compared to the best fit point) of $50 \%$.

The allowed fit region for CMS can be compared to the recently presented public results \cite{Wedtalk}, which restrict the fit to the region $c>0$ (physically different from the region $c<0$) 
\footnote{We can always set $a>0$ by a redefinition of the $h$ field. The sign of $c$ could in principle be changed also by rotating the fermion fields, but this would affect in the same way the fermion mass so that the sign of $c/m_f$ is in fact fixed and physically meaningful.}. As the absolute minimum of the $\chi^2$ lies in fact in the discarded region, the shape and size of the $68\%$ and $95\%$ C.L. regions presented by CMS differ from the ones
that we obtain in figure~\ref{individualexp}. If we also restrict 
our parameter space to positive $c$ we have checked that we get excellent agreement with the CMS result. Note however that there is no valid reason to discard a priori the negative $c$ region which offers in fact the interesting possibility of giving
a good fit to the data by an enhancement of the $h\gamma\gamma$ coupling through constructive interference of the top and $W$ loops.

\subsection{BSM Implications}\label{BSM}

\subsubsection{Implications for an invisible width}

The results of the last section can be interpreted as (partial) evidence in support of the SM Higgs hypothesis. Assuming then that the observed boson is the SM Higgs, we can study possible deviations of its properties due to BSM effects. The simplest extension of the SM in terms of new parameters is perhaps the case where one only introduces a Higgs invisible branching ratio,  ${\rm Br}_{inv}$. Such an extension is common in many BSM scenarios, {\it e.g.} when new physics couples through the Higgs portal,
and new states exist that are uncharged under the SM group. Fitting to  ${\rm Br}_{inv}$ allows one to fit to the global combined signal strength values supplied by the experimental collaborations. The values we use
to perform the fit are given in Table \ref{table:combineddata}.

\begin{figure}[tb]
\includegraphics[width=0.32\textwidth]{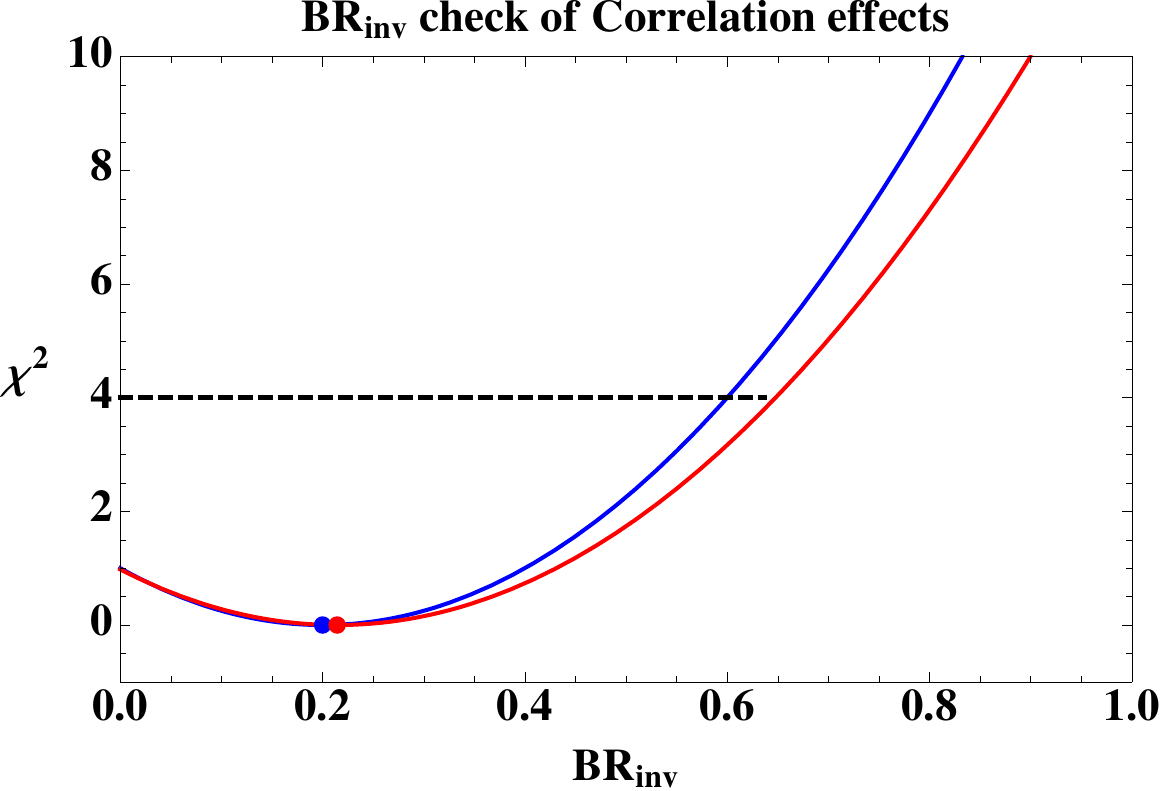}
\includegraphics[width=0.32\textwidth]{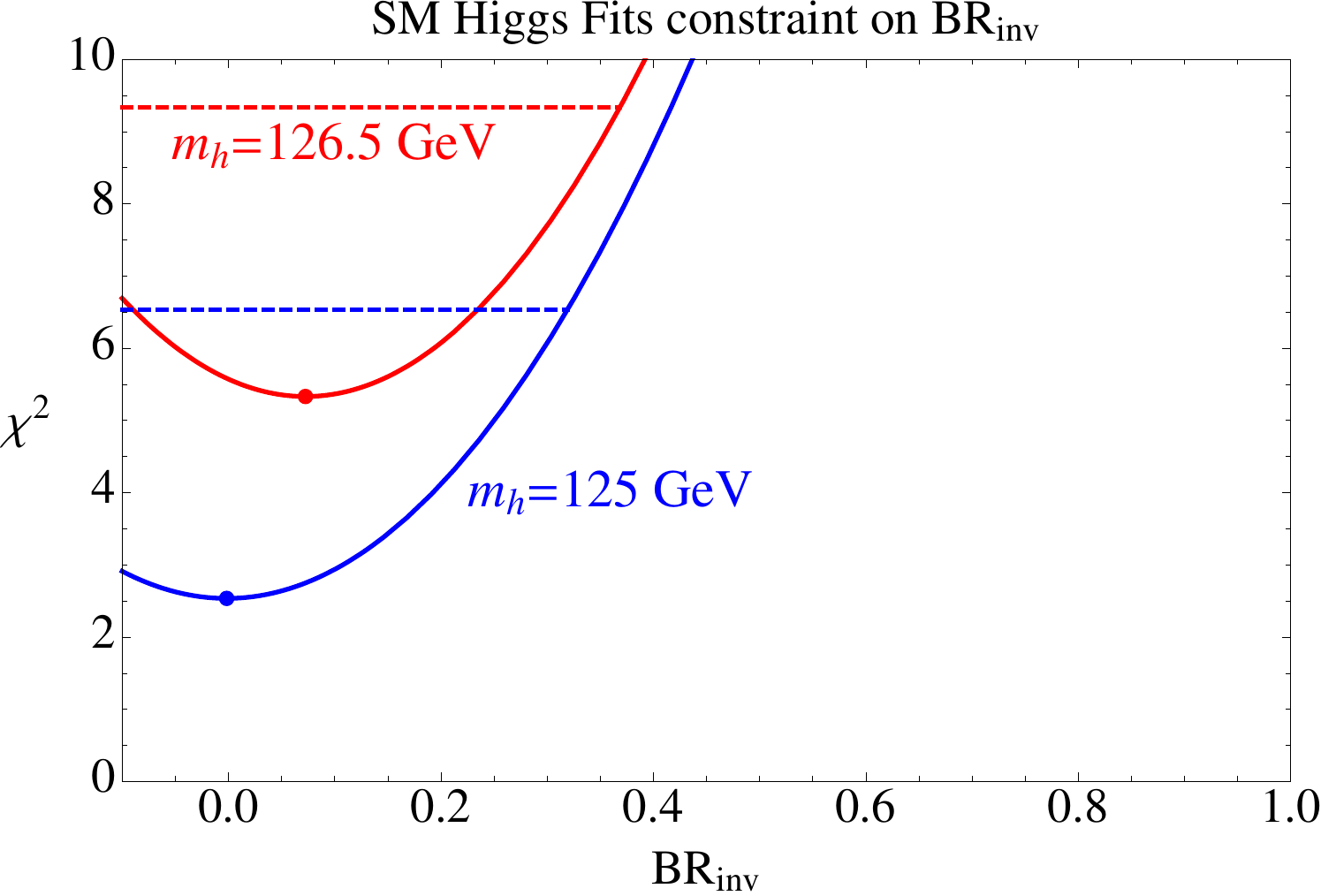}
\includegraphics[width=0.32\textwidth]{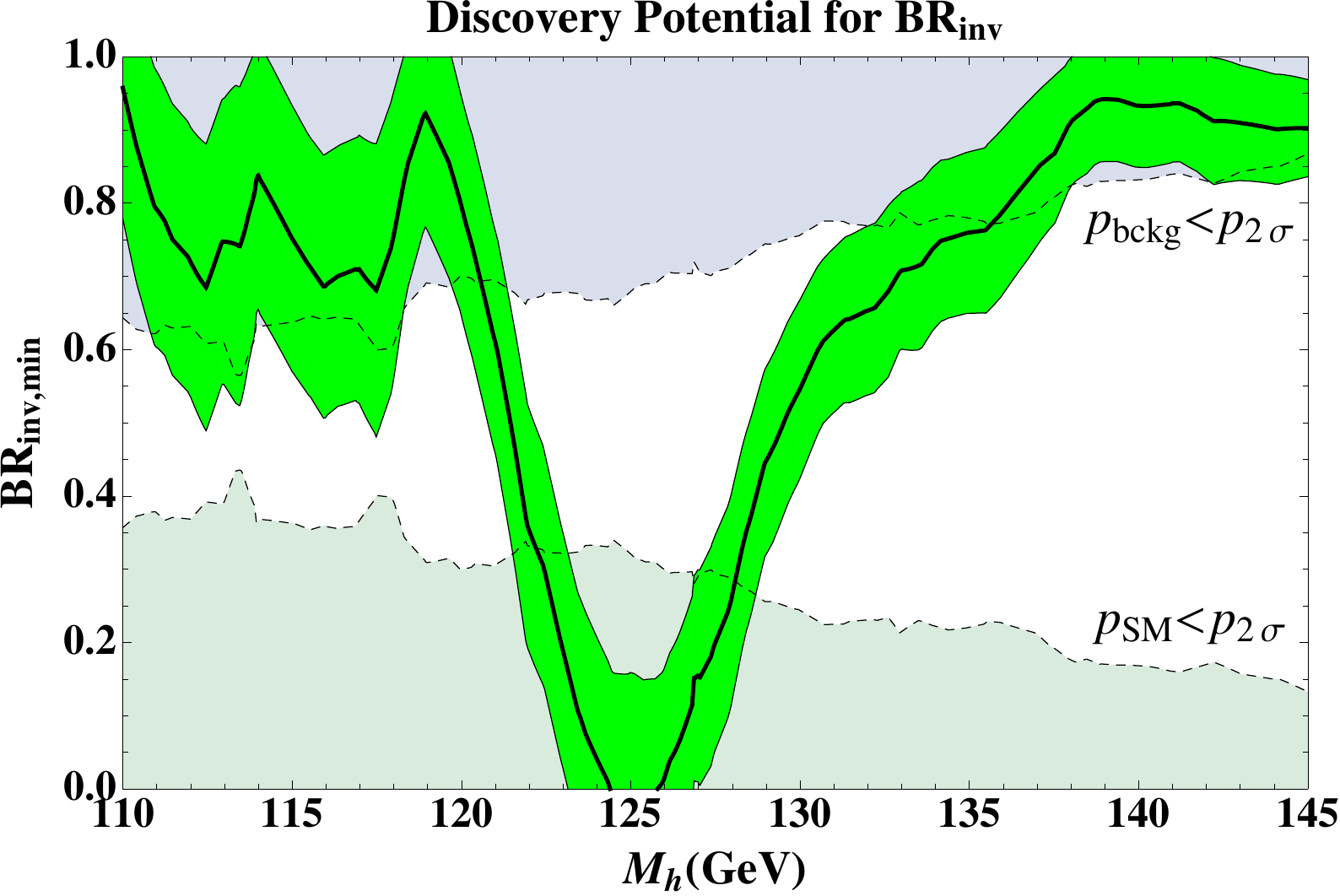}
\caption{\it Global fit to ${\rm Br}_{inv}$ for the SM Higgs using only CMS data (left figure) for $m_h = 125 \, {\rm GeV}$ with two methods as a check of our fit procedure. Blue curve - global signal strength based fit, Red curve - individual channel fit. See text for further explanation. The middle figure shows
the $\chi^2$ distribution developed from the combined best fit $\hat{\mu}_c$ supplied by the four experiments, including the 7 and 8 TeV LHC results for two mass values.
In the right figure we show the discovery potential for ${\rm Br}_{inv}$, updating a result from the analysis in Ref.~\cite{Espinosa:2012vu} with the new global signal strength data. }
\label{Fig.globalSM1}
\end{figure}

\begin{table}[h] 
\setlength{\tabcolsep}{5pt}
\center
\begin{tabular}{c|c|c|c|c} 
\hline \hline 
Experiments & $\hat\mu_c$, \, $m_h =125$ & $\sigma_c$,\, $m_h = 125$ & $\hat\mu_c$,\,  $m_h = 126.5$ & $\sigma_c$,\,  $m_h = 126.5$
\\
\hline
${\rm CMS}$ [7\&8 \,TeV] \, \cite{Wedtalk} & $0.80$ & $0.20$ & $0.67$ & $0.19$
\\
${\rm ATLAS}$ [7\&8 \,TeV] \, \cite{Wedtalk} & $1.12$ & $0.27$ & $1.24$ & $0.26$
\\
${\rm ATLAS}$ [7\&8 \,TeV] (\& $\mu_{WW}$) \, \cite{atlaslatest} & $1.32$ & $0.29$ & $1.37$ & $0.27$
\\
${\rm CDF \& D 0\! \! \! /}$  \, \cite{tevatronupdate}& $1.35$ & $0.59$ & $1.38$ & $0.60$
\\
\hline \hline
\end{tabular}
\caption{\it Combined signal strengths $\hat{\mu}_c$ and errors $\sigma_c$ from ATLAS, CMS and the Tevatron collaborations. Here we quote $\pm$ symmetric $1 \sigma$ errors.}
\label{table:combineddata} \vspace{-0.35cm}
\end{table}

The left plot of Fig.\ref{Fig.globalSM1} shows the result of extracting ${\rm Br}_{inv}$ using our fit approach using two different methods.
This allows an important cross check of our procedure and results. We have employed in our fit two approximations that require further justification. Using the assumption of Gaussian PDF's, we have extracted the $8 \, {\rm TeV}$ data from the known $7 \, {\rm TeV}$ data and the released $7+8 \, {\rm TeV}$ data. In doing so we
have neglected (unsupplied) correlations. Further, in performing our fits we have neglected correlation coefficients for the $\mu_i$. The left plot of Fig.\ref{Fig.globalSM1} shows in blue
the result of fitting directly to the combined CMS data  - which do take into account correlations. In red we show, for comparison, the fit results when we use our procedure to extract the $8 \, {\rm TeV}$ data, and fit to the individual $\mu_i$'s. The best fit point and the $95 \%$ C.L. regions are in good agreement, supporting the estimated accuracy of our results of $\sim 5-10 \%$.
 
The middle plot of Fig.\ref{Fig.globalSM1} shows instead the resulting $\chi^2$ distributions for  ${\rm Br}_{inv}$ for two different mass values  $m_h = 125,126.5 \, {\rm GeV}$ when we combine the results from the three experiments and
 fit to the supplied signal strength parameters $\hat\mu_c, \sigma_c$.  This is the most accurate analysis on 
${\rm Br}_{inv}$ that we can perform with the released data (more accurate than a fit on the individual channels). This is primarily due to the experimental correlation effects that are incorporated in the $\hat{\mu}_c$, and larger number of channels that are incorporated in the experimental likelihoods used to construct the combined signal strength parameter used. We find the $95 \%$ C.L. regions ${\rm Br}_{inv} < 0.37 (0.40)$ for $m_h = 125 (126.5) \,{\rm GeV}$. These limits can be used to constrain many models that predict an invisible Higgs width. For recent related work, see Ref.~\cite{Espinosa:2012vu,Giardino:2012ww,Frigerio:2012uc}.
 
The right plot  of Fig.\ref{Fig.globalSM1} (which updates a similar analysis in Ref.~\cite{Espinosa:2012vu})  shows the current status of the quantity $(1-\hat\mu_c)$ (with one sigma error band), interpreted as an invisible Higgs branching ratio, as a function of the Higgs mass. This result is also based on the combined signal strengths supplied by the experimental collaborations. With the current errors, no significant statement can be made about the possible nonzero central value of  ${\rm Br}_{inv}$. The shaded areas indicate the 2$\sigma$ range of possible fluctuations (of the background in the upper region at large  ${\rm Br}_{inv}$; of the SM Higgs signal in the lower region at small  ${\rm Br}_{inv}$) that could be miss-interpreted as an invisible Higgs width. This plot shows that the current nonzero central values of ${\rm Br}_{inv}$ at $m_h\sim 125 \, {\rm GeV}$ (consistent with the fit results) are perfectly compatible with downward fluctuations of the SM Higgs signal.

The main difference between the results presented in Fig.\ref{Fig.globalSM1} with respect to the first version of this paper comes from the use of the ATLAS combined signal strength reported in Ref.~\cite{atlaslatest}, that incorporates now  the large 8 ${\rm Tev}$ ATLAS WW ~\cite{ATLAS-CONF-2012-098} signal strength $ \mu_{WW}(m_h =125) = 2.1^{+ 0.8}_{- 0.7}$. One finds the $95 \%$ C.L. limits ${\rm Br}_{inv} < 0.32\, (0.37)$ for $m_h = 125 (126.5) \,{\rm GeV}$ when this signal strength is added to the global data set. 

%
%
\subsubsection{Implications for $c_g,c_{\gamma}$}

One can also infer the current experimental bounds on the BSM parameters $c_g, c_{\gamma}$.
We expect that  these operators arise at the loop level, so we rescale the Wilson coefficients as $c_j = \tilde{c}_j/(16 \pi^2)$ for $j = g,\gamma$.
Using the results of Ref.~\cite{Manohar:2006gz}, the effects of these operators are incorporated as rescaling factors used in the fit and given by
\bea\label{higher-d.effect}
R_{g} \equiv \frac{\sigma_{gg \rightarrow h}}{ \sigma^{SM}_{gg \rightarrow h}} &\approx& \left| 1 - \frac{1}{0.75 c_t -(0.05 +0.07 i) c_b} \frac{v^2 \, \tilde{c}_{g}}{\Lambda^2}\right|^2,  \\
R_{\gamma} \equiv \frac{\Gamma_{h \rightarrow \gamma \, \gamma}}{\Gamma^{SM}_{h \rightarrow \gamma \, \gamma} } &\approx&  \left| 1 + \frac{1}{4 (2.07 a - 0.44 c_t + (0.01 + 0.01 \, i) c_b)} \frac{v^2 \, \tilde{c}_{\gamma}}{\Lambda^2}\right|^2.  
\eea
Here we have used $m_t = 172.5 \, {\rm GeV}$, $m_b = 4.75 \, {\rm GeV}$, $m_h = 125 \, {\rm GeV}$ and $\alpha_s(172.5\ {\rm GeV}) = 0.107995$. 
When the Higgs has SM renormalizable couplings, then $a = c_t = c_b = 1$. We have retained the two-loop QCD correction to the SM matching of the $h \, G^A_{\mu\, \nu} G^{A \, \mu \, \nu} $ operator in the $m_t \rightarrow \infty$ limit in these numerical coefficients. 
The operators in $\mathcal{L}^5_{HD}$ also affect ${\rm Br}(h \rightarrow \gamma \, Z)$, but the effects in our fit can be neglected. 
See Ref.~\cite{Espinosa:2012vu} for further details and discussion on our fitting procedure to these higher dimensional operators.  

\begin{figure}[tb]
\includegraphics[width=0.33\textwidth]{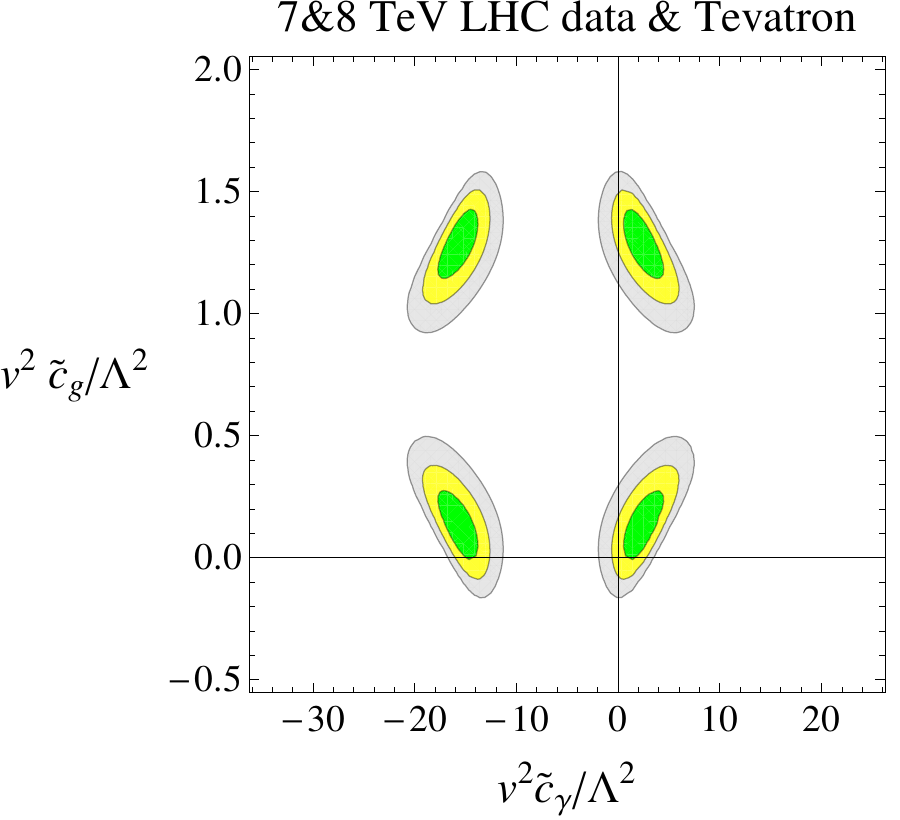}
\includegraphics[width=0.33\textwidth]{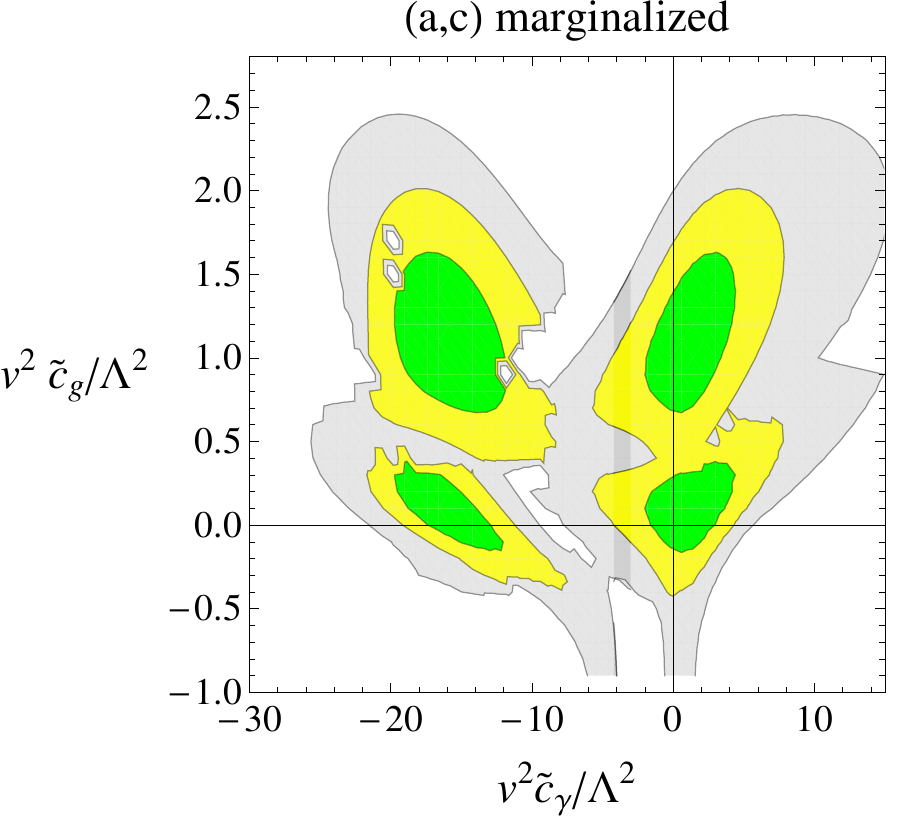}
\includegraphics[width=0.32\textwidth]{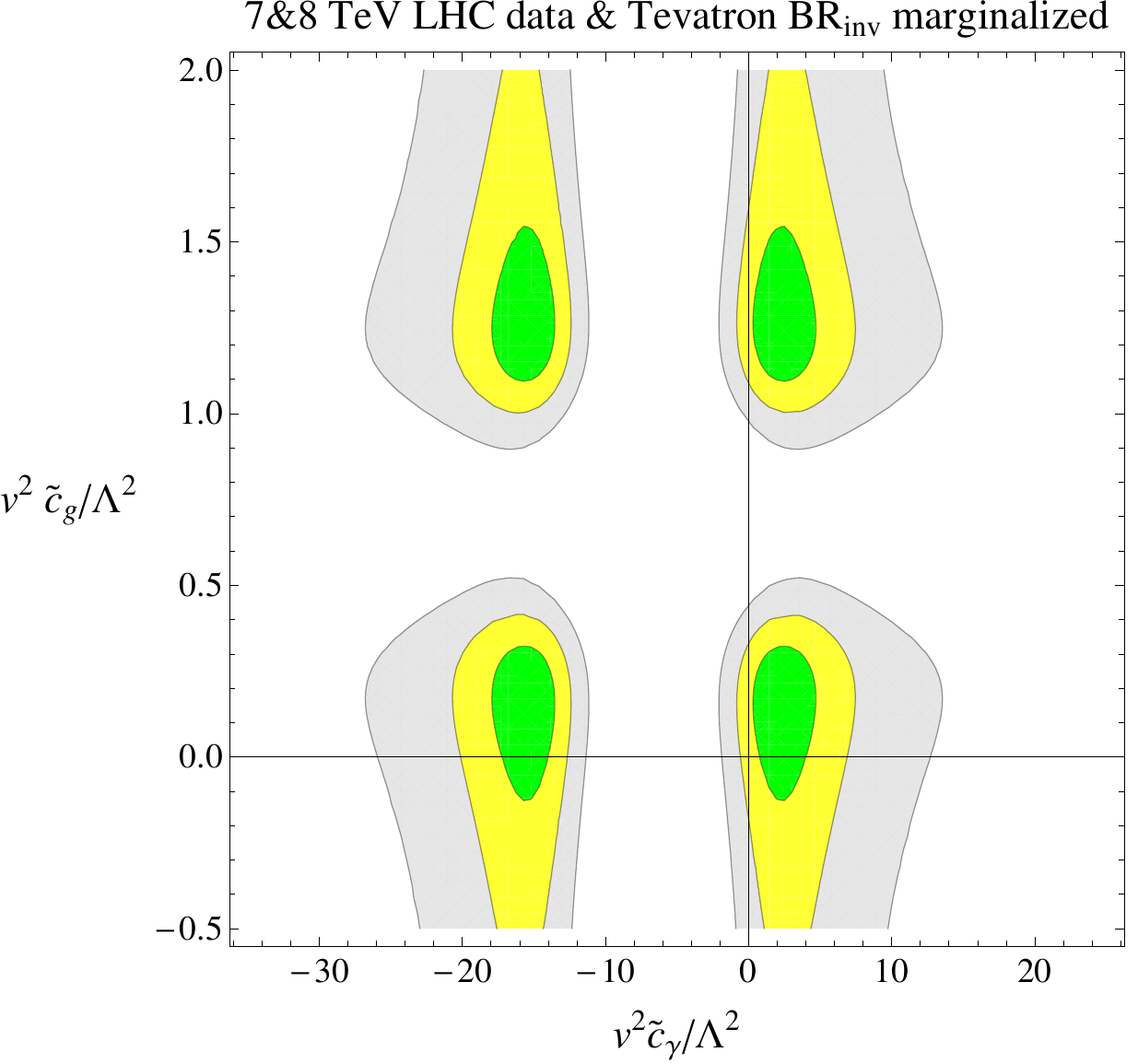}
\caption{(Left) Results of fitting to $c_g, c_{\gamma}$ when the SM is assumed. Note that in these results we have assumed that $c_g,c_\gamma$ are real. (Middle) Results of fitting to $c_g, c_{\gamma},a,c$ and marginalizing over
(a,c) subject to the constraint $a>0$ and $0 < c < 3$. (Right) Results of fitting to $c_g, c_{\gamma},{\rm BR}_{inv}$ and marginalizing over ${\rm BR}_{inv}$.}\label{higherdfigure}
\end{figure}

We show in Figure \ref{higherdfigure} the results of fitting to $c_g, c_{\gamma}$ using the current dataset when one assumes the SM values $a=c=1$ (left); when one
fits to $c_g, c_{\gamma},a,c$ and subsequently marginalizes over $a,c$ (middle); and finally, when one fits to $c_g, c_{\gamma},{\rm BR}_{inv}$ and marginalizes over ${\rm BR}_{inv}$.
These results summarize the preference in the current dataset for including higher dimensional operators when the scalar is not assumed to be the Higgs.

\begin{figure}[tb]
\includegraphics[width=0.4\textwidth]{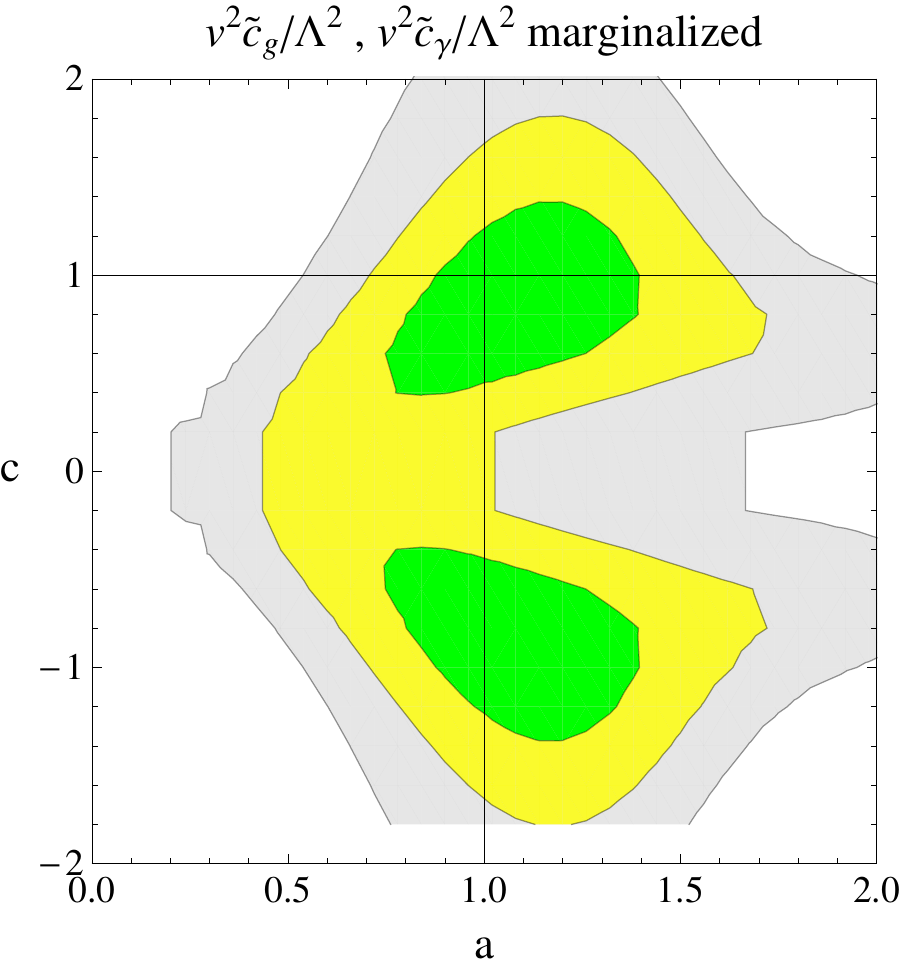}
\includegraphics[width=0.4\textwidth]{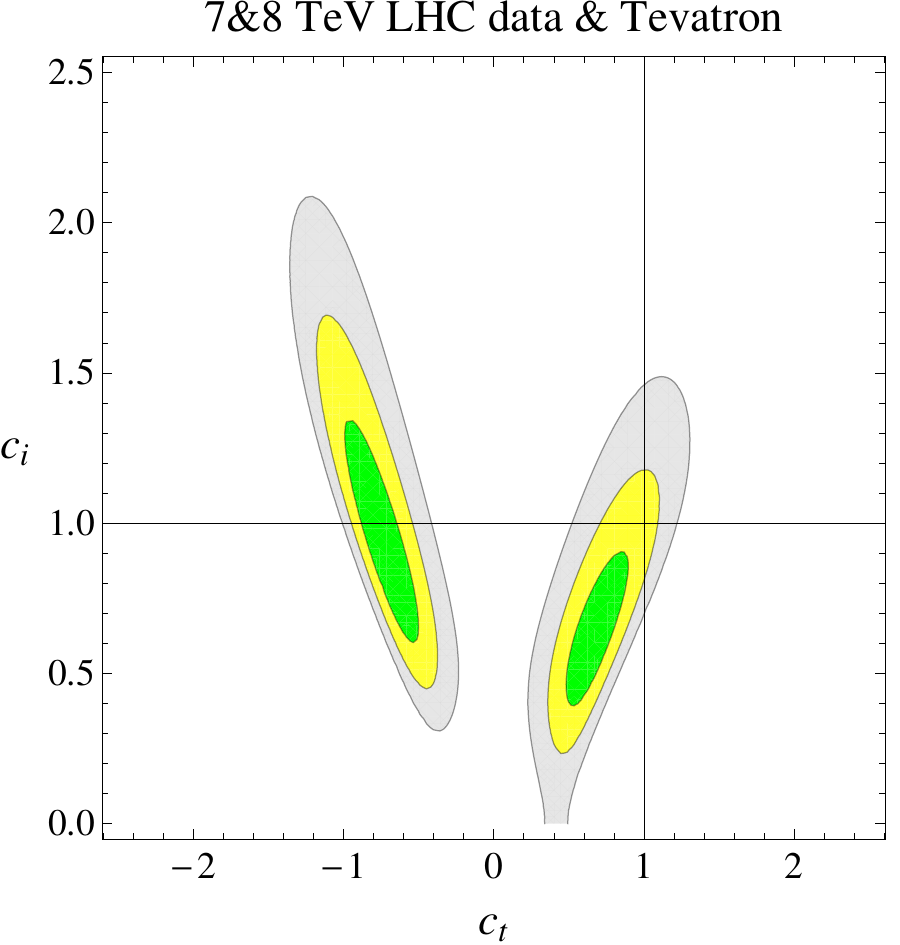}
\caption{(Left) Results of fitting to $c_g, c_{\gamma},a,c$ and marginalizing over the higher dimensional operators.
(Right) Allowed space when $c_t$ is varied independently. See text for more details.}\label{moremarg}
\end{figure}

We also show in  Fig.~\ref{moremarg} (left) the fit to $c_g, c_{\gamma},a,c$, marginalizing over the higher dimensional operators. This case shows the preference for $a,c$
even when more massive states are integrated out of a composite scalar theory (for example) contributing to $c_g, c_{\gamma}$. We see that the SM hypothesis is significantly improved in its 
consistency with the dataset in the context of NP of this form. This is particularly relevant as it directly shows what the data tells about $a$, the key parameter that probes the involvement of $h$ in electroweak symmetry breaking.  A final plot in Fig.~\ref{moremarg} (right)
shows the allowed space when $c_t$ is varied independently from the remaining fermion couplings which are treated with a universal rescaling.

The fit to higher dimensional operators shows a preference for a BSM contribution to $c_\gamma$, which deserves some comment. First, one can study the model independence of this preference for an enhancement of $c_\gamma$
by constructing a one dimensional $\chi^2$ distribution, marginalizing over the unknown operator $c_g$ when the SM values $a=c=1$  are assumed, or over $(c_g,a,c)$ when one is considering 
non-SM scalar scenarios. Performing this exercise we find that 
a preference for $c_\gamma \neq 0$ only exists in certain cases where implicit UV assumptions are adopted due to the parameters used to fit the data.  One can make a number of observations regarding enhanced
$h \rightarrow \gamma \gamma$ event rates. The coupling to $F_{\mu \, \nu}$ can come about due to $c_{WB}$.
If this is the case, one can bound the allowed enhancement of $\mu_{\gamma \, \gamma}$ due to related EWPD constraints. Using the relation
\cite{Grinstein:1991cd,Han:2004az,Manohar:2006gz}
\bea
\frac{v^2}{\Lambda^2} \tilde{c}_{WB} =  - 2 \, \pi \, S,
\eea
one finds the current experimental constraint $|v^2 \, \tilde{c}_{WB}/\Lambda^2| \lesssim N \, 0.63$ for an $N \sigma$ deviation considered in the EWPD
parameter $S$, assuming $m_h = 125 \, {\rm GeV}$. Using $\mu_{\gamma\, \gamma} \approx R_g \, R_{\gamma}$, this leads to the bound $\mu_{\gamma\, \gamma} < 1.2 \, (1.4)$ for a one (two) sigma deviation in the $S$ parameter when $c_g = 0$. 
Restricting the unknown higher dimensional operators to have a global $\chi^2$ measure in the $2 \sigma$ allowed region ($\Delta \chi^2(c_g,c_\gamma) < 6.18$),
one can maximize the enhancement of $\mu_{\gamma \gamma}$ considering the related constraints on $c_{WB}$. One finds that an enhancement of 1.4
is possible when a one sigma deviation in the $S$ parameter is also allowed. The $c_{\gamma}$ Wilson coefficient could also come about due to $c_B,c_W$ or a combination of the two.  In this case, the EWPD constraint does not directly apply, and an enhancement of $\mu_{\gamma\, \gamma}$ 
by a factor of $\lesssim 1.7$ is still allowed when maximizing the contribution subject to the constraint $\Delta \chi^2(c_g,c_\gamma) < 2.3$. 

This analysis has assumed no relationship between the Wilson coefficients  $c_\gamma,c_g$, which is fixed in any particular UV completion.
It is interesting to note that in general when integrating out a single BSM field one expects a strong relationship between the Wilson coefficients, with identical loop functions in many cases,
the only differences in the matching onto the Wilson coefficients is due to the relative charges of the new states under the SM subgroups ${\rm SU}(3)_c$ and ${\rm U}(1)_{\rm em}$.
 
\section{Studies of Consistency and Tension within the Dataset}\label{tension}

\begin{figure}
\includegraphics[width=0.45\textwidth]{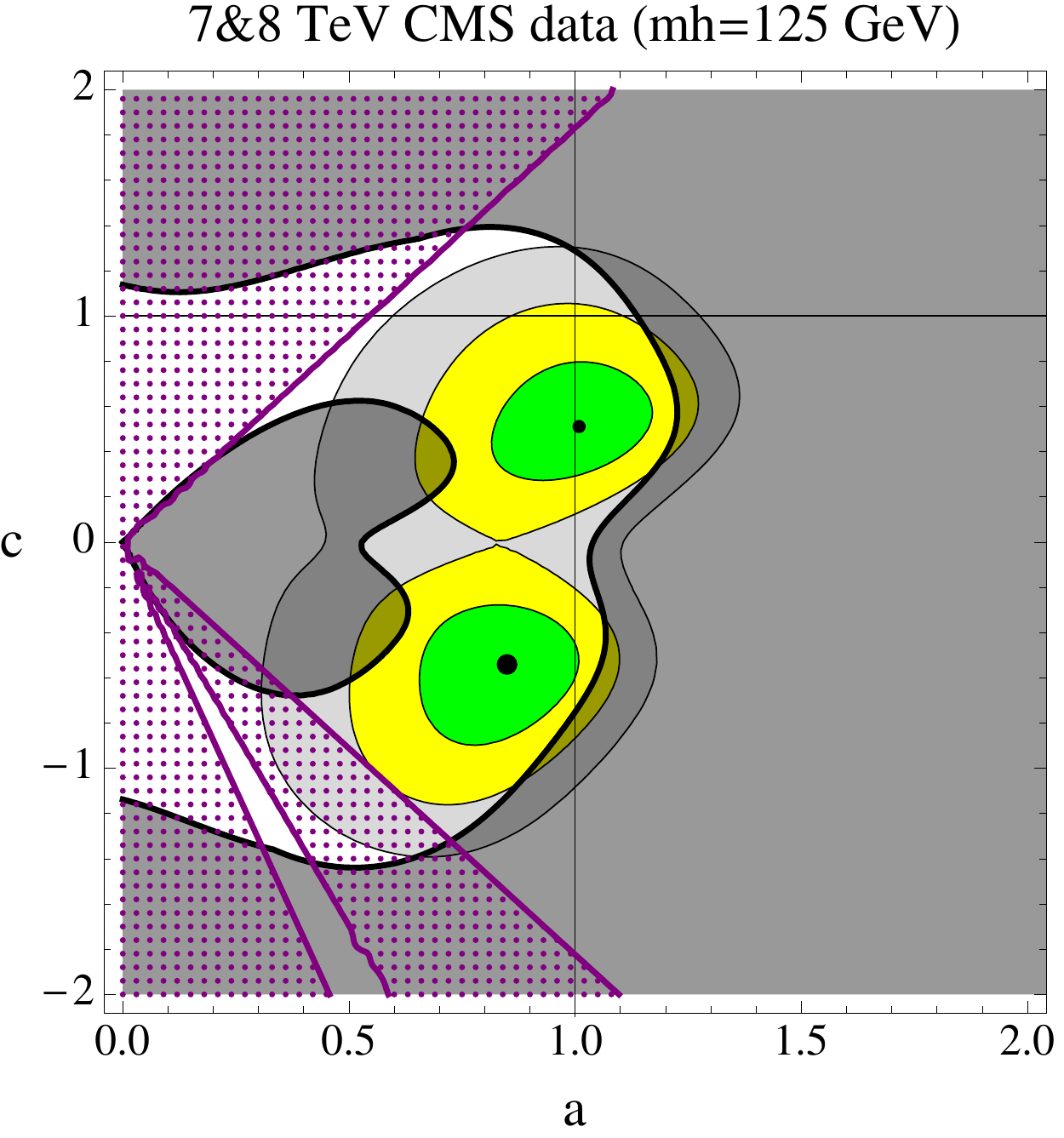}
\includegraphics[width=0.45\textwidth]{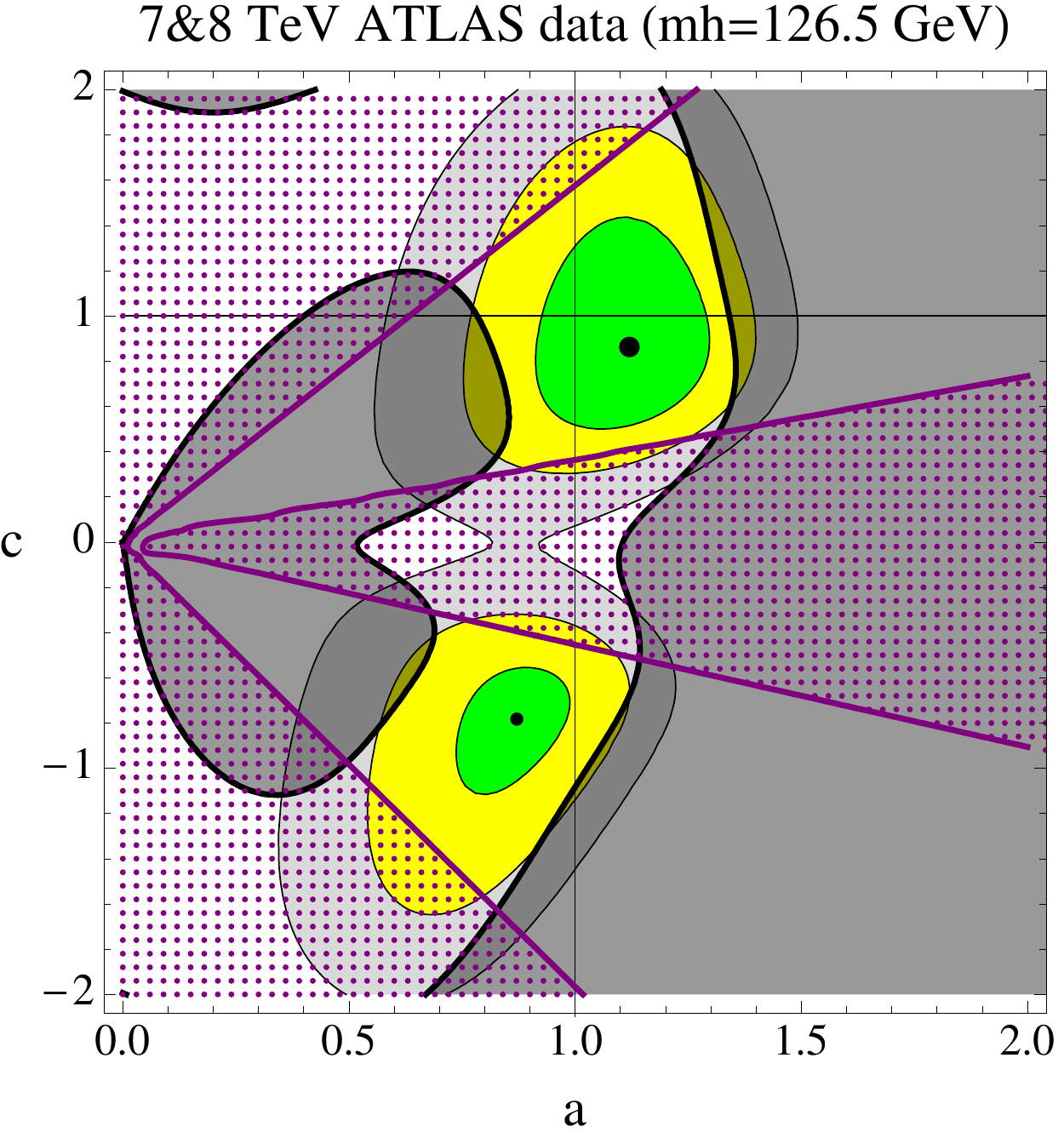}
\includegraphics[width=0.45\textwidth]{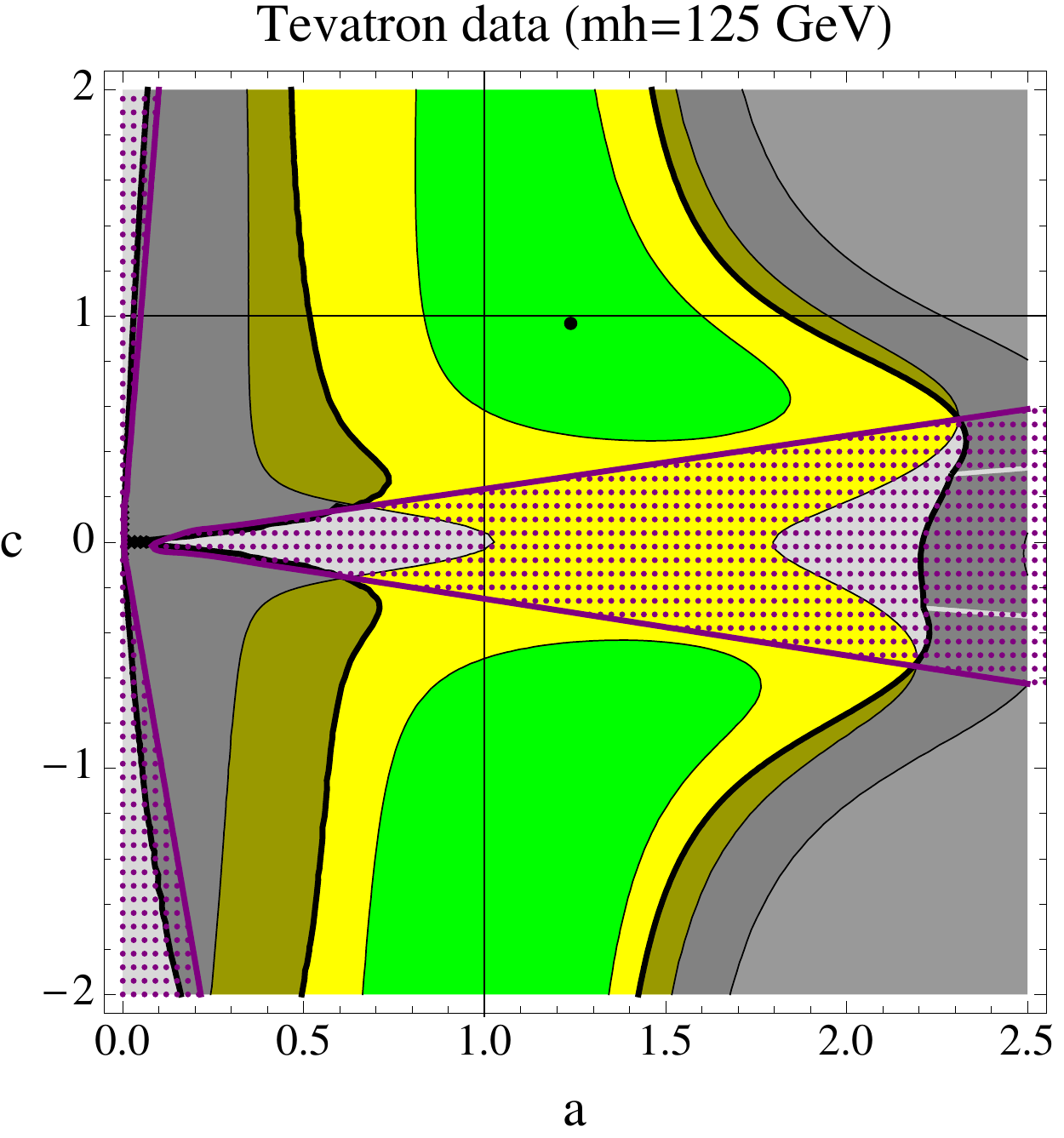}
\includegraphics[width=0.45\textwidth]{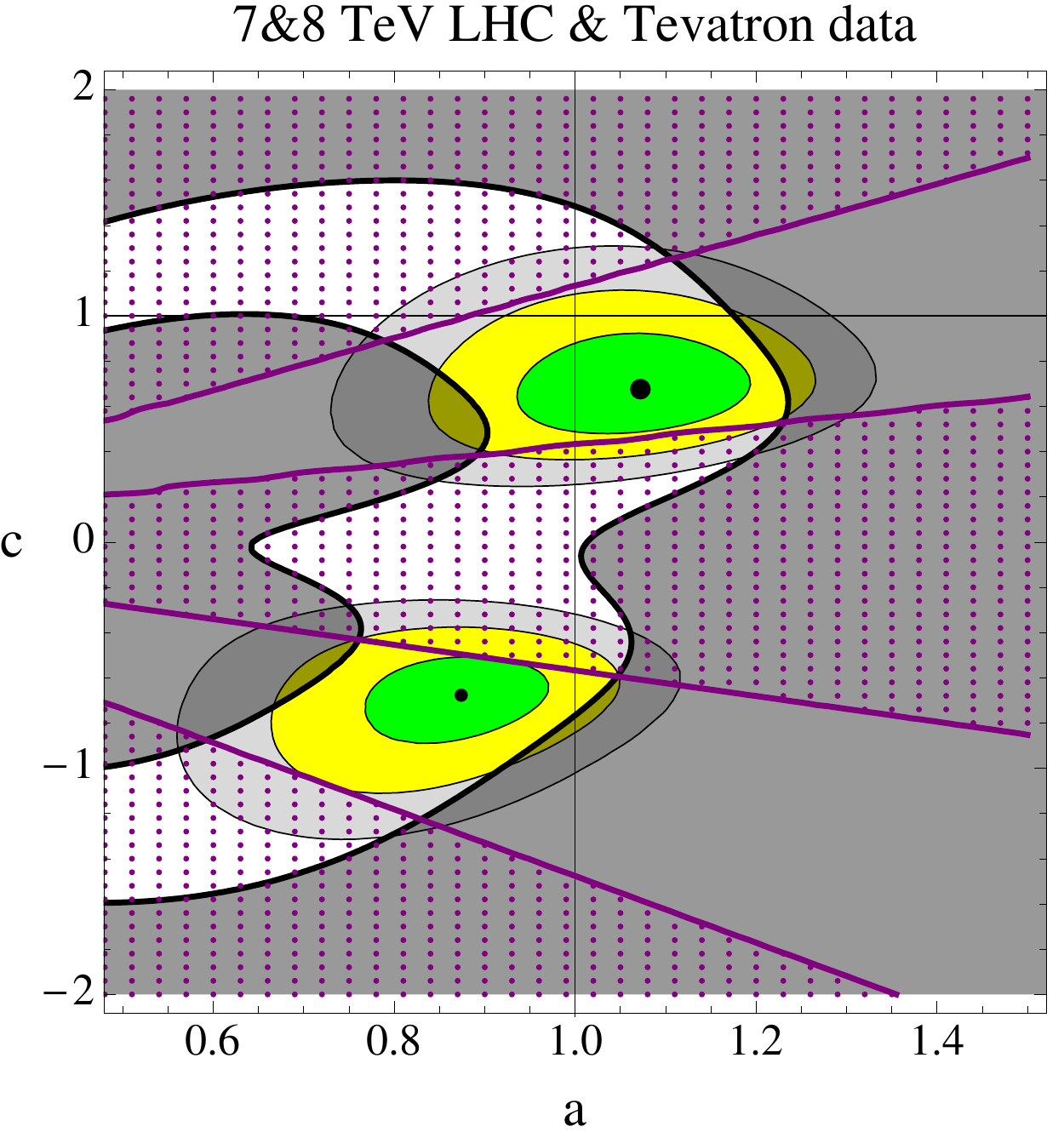}
\caption{Exclusion limits on the ${\rm SM}(a,c)$ parameter space derived from a)
95\% C.L. upper limits on the signal strength parameter $\hat\mu<\mu_{upL}$ (gray shaded regions beyond the black lines towards the right), b) 95\% C.L. lower limits $\hat\mu>\mu_{dwL}$ (gray shaded regions enclosed by the black lines close to the origin) and c) "tension limits"  (purple dotted regions delimited by purple straight lines) from the presence of search channels inconsistent with the rest of the dataset  
at 2 $\sigma$. The most powerful channels in setting such limits are listed in the text.
For comparison, the best-fit regions discussed in previous sections are also shown. The fermiophobic scenario along $c=0$ is excluded.
}\label{tenslim}
\end{figure}

The consistency between the search results ($\hat\mu_i\pm\sigma_i$) from
different channels can be quantified as follows. For each channel $i$ we
construct its Gaussian approximation to the pdf of the signal strength,
${\rm pdf}_i(\mu)={\rm pdf}(\mu,\hat\mu_i,\sigma_i)$, which we can contrast with the full
combined PDF (either in a given experiment or for the overall combination of
all data), ${\rm PDF}(\mu)=pdf(\mu,\hat\mu_c,\sigma_c)$. For each single channel we calculate its $p$-value with respect to
the global combined PDF, {\it i.e.}, $p_{ic}$ is the $p$-value for $\hat\mu_i$ assuming the full ${\rm PDF}(\mu)$. We can also define the $p$-value of the global $\hat\mu_c$ with respect to the individual ${\rm pdf}_i(\mu)$, which we will denote by $p_{ci}$. In order to treat properly the $p$-values for cases with $\hat\mu_i<0$, we will normalize the pdf's in this section
so that they give 1 when integrated over the whole interval $\mu\in(-\infty,\infty)$. We then have, for $\hat\mu_i<\hat\mu_c$,
\bea
p_{ic} & \equiv &
 \int_{-\infty}^{\hat\mu_i}{\rm PDF}(\mu)d\mu = \frac{1}{2}{\rm erfc}\left(\frac{\hat\mu_c-\hat\mu_i}{\sqrt{2}\sigma_c}\right)\ , \nonumber\\
p_{ci}&\equiv &\int_{\hat\mu_c}^{\infty}{\rm pdf}_i(\mu)d\mu= \frac{1}{2}{\rm erfc}\left(\frac{\hat\mu_c-\hat\mu_i}{\sqrt{2}\sigma_i}\right)\ ,\;\;\;\;
\eea
and, for $\hat\mu_i>\hat\mu_c$, 
\bea
p_{ic} & \equiv &
 \int_{\hat\mu_i}^{\infty}{\rm PDF}(\mu)d\mu = \frac{1}{2}{\rm erfc}\left(\frac{\hat\mu_i-\hat\mu_c}{\sqrt{2}\sigma_c}\right)\ , \nonumber\\
p_{ci}&\equiv&\int_{-\infty}^{\hat\mu_c}{\rm pdf}_i(\mu)d\mu=  \frac{1}{2}{\rm erfc}\left(\frac{\hat\mu_i-\hat\mu_c}{\sqrt{2}\sigma_i}\right)\ ,\;\;\;\;
\eea
where ${\rm erfc}$ is the complementary error function, ${\rm erfc}(z)=1-{\rm erf}(z)$.
We will say channel $i$ is in tension with the rest of the data if $p_{ic}$ and $p_{ci}$ are both very small. For a given critical $p$-value $p_N$, corresponding to $N$ standard deviations, channel $i$ is not consistent with the combined dataset at $(1-p_N)$ C.L. if $p_{ic},p_{ci}<p_N$ and the model can be
excluded based on that disagreement. The consistency condition reduces simply to
\be
|\hat\mu_i-\hat\mu_c|<N\ {\rm max}[\sigma_i,\sigma_c]\ .
\ee
We will choose $N=2$ in our discussion.

When this test for consistency is applied to the ($\hat\mu_i\pm\sigma_i$)
dataset, interpreted as coming from a SM Higgs signal, we find tension
at this $2\sigma$ level for two ATLAS $\gamma\gamma$ channels at 7 TeV, those labelled URhPTh and CChPTt (the two outliers easily identified in Fig.~\ref{Fig.data}). In extensions of the SM, like the two-parameter scenario ${\rm SM}(a,c)$ we have discussed in previous sections,
the rescaling of the different channels (which affects $\hat\mu_i\pm\sigma_i$ as explained in section \ref{mudata}) can introduce very significant distortions
in the pdfs and cause too large tensions for some other channels. Such regions of
parameter space can therefore be excluded on this basis.
Figure~\ref{tenslim} illustrates this by showing, on addition to the best fit regions and the 95\% C.L. exclusion regions derived from upper and lower limits on the signal strength,  the regions in $(a,c)$ space that would be excluded due to more than 2-$\sigma$ tension in some search
channel (purple shaded regions delimited by straight lines).
We show such limits both experiment by experiment and for the combined result.
Typically, in the excluded regions several channels at the same time
cause the exclusion. We show in each case all the region that is excluded by
at least one channel (with the exception of the ATLAS case, see below). In each case, the channels that have bigger exclusion power are the following. For Tevatron, the $b\bar{b}$ channel; for CMS, at 7 TeV, $\gamma\gamma_{jj}$ and at 8 TeV, $\gamma\gamma_3$ and $\tau\tau$; for ATLAS, the two $\gamma\gamma$ channels mentioned above,  URhPTh and CChPTt at 7 TeV, cause $2\sigma$ tension in all the region of parameter space shown and we do not mark this area. Besides these channels, there is $2 \, \sigma$ tension also from $\gamma\gamma_{CClPTt}$ at 7 TeV and from $WW\to ll\nu\nu$ at 8 TeV.
The fact that the tension limits are straight lines passing through the origin is due to the fact that any common rescaling of $a$ and $c$ (that leaves $c/a$ invariant) also leaves invariant the tension associated to any channel as the latter are functions of ratios $(\mu_i-\mu_c)/\sigma_{i,c}$, and such ratios are also functions of $c/a$.
We see from Figure~\ref{tenslim}, that fermiophobic scenarios, along the
axis $c=0$, are excluded at this level of confidence.

This channel-tension analysis is clearly related to the $\chi^2$ study we
have also performed and the tension exclusion limits tend to exclude regions
of parameter space that give a bad fit to the data. However, this approach
seems to be more powerful in being able to exclude definitely some models. We
also see that there are regions of parameter space that are not excluded by
the conventional upper (or lower) limits imposed on the signal strength
parameter,
yet can be excluded by the tension exclusion limit. This is simply due to the
fact that the dataset can contain two widely separate channels and still give
a combined PDF that respects the upper limit, which is not able to probe in
such cases the internal inconsistency of the individual channels. We conclude
that this type of analysis offers a complementary tool in testing model
performance.

\section{Conclusions}\label{concl}

We have studied the evidence of a scalar field that has been  discovered by the LHC collaborations using an effective field theory
framework. We have discerned what can be inferred about the properties of the scalar field at this time using joint $\chi^2$ fits to available datasets.
We have also developed and applied new techniques that allow one to exclude model classes that are in tension with the data through violating
upper or lower C.L. bounds, or through introducing excessive tension into the signal strength parameter datasets. At this time, according to our fit method,
and using publicly available data, the SM Higgs hypothesis
is consistent with the global dataset compared to the best fit point at the level of $\lesssim 2 \sigma$.

\appendix

\section{Data Used}

Our approach to the presented data is as follows. As the relative weights of the various contributions to the inclusive Higgs production processes depend on the operating energy of the LHC, we need separate information on $7$ and $8 \, {\rm TeV}$ data to perform our fit as the relative fraction of events at each operating energy depends on the unknown parameters $a,c,c_g,c_\gamma$.\footnote{Fits to ${\rm BR}_{\rm inv}$ do not suffer this problem as this parameter affects universally all channels.} 
Most of the public results \cite{Wedtalk} are now presented separately in $7,8 \, {\rm TeV}$ signal strengths. When this information is available we use directly this data.
 
\begin{figure}[ht]
\includegraphics[width=1\textwidth]{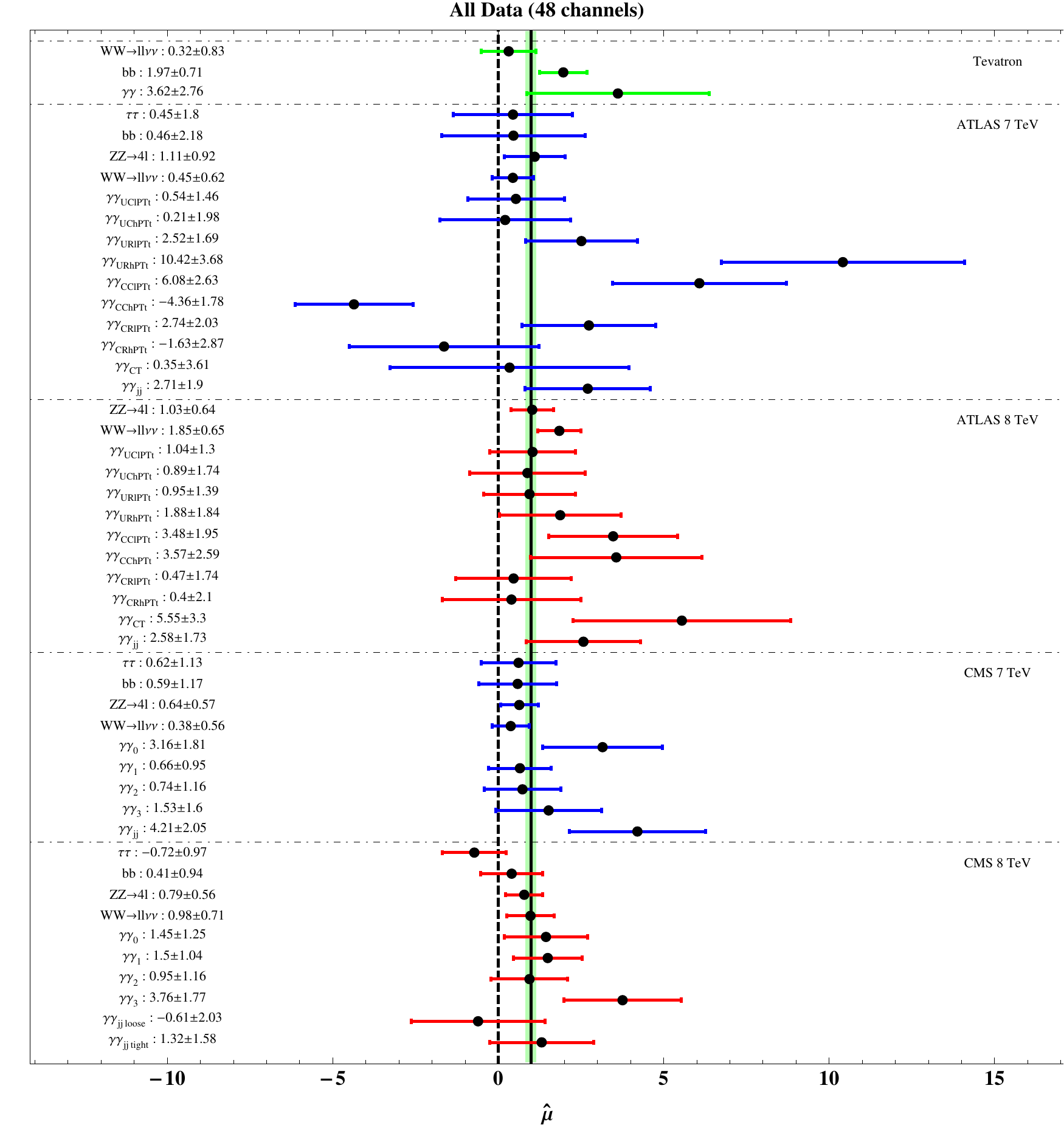}
\caption{\it Pictorial presentation of the data used in the fits to sub-channels. Blue: reported data at $7$ TeV. Red: reported $8\, {\rm TeV}$ data, or reconstructed $8 \, {\rm TeV}$ data.}
\label{Fig.data}
\end{figure}

For those cases in which only the combined $7+8$ TeV signal strengths are available (currently CMS combined $\hat{\mu}_{\tau \, \tau}$, $\hat{\mu}_{WW}$ used in the fit)
in addition to the $7$ TeV results, we make use of Eq.~(\ref{csigma}) to reconstruct the unreported $8 \, {\rm TeV}$ data. Note that this relies in the use of Gaussian approximations to the PDF's to describe the data (and signal strengths) and should be increasingly accurate as the total number of events increases. This can be done without knowing directly the experimental likelihood function in the limit that correlations are neglected, which is an assumption we are already forced to adopt as this information 
is not supplied by the experimental collaborations. We show in Fig.~\ref{Fig.data} the resulting reconstructed data. 

An interesting check of our approach is to use a subset of the provided subclass signal strengths to reproduce a reported combined signal strength.
This exercise can be carried out, for example, on the supplied vector boson fusion tagging $b \, \bar{b}$ signal strength and $t \, \bar{t} \, h$ signal strength (that uses $h \rightarrow b \, \bar{b}$)
and comparing to the presented combined $h \rightarrow b \, \bar{b}$ signal strength. We have carried out this procedure for the presented CMS data and find good agreement with the reported results,
within our estimate of $5-10 \%$ error introduced due to a lack of correlations.

We have updated our fit from
v1 of this paper to include the following information that has been released since the first version appeared on the archive. The full subclasses of $\gamma \, \gamma$ events are now available from ATLAS and CMS at  $7,8 \, {\rm TeV}$.
Also, the production channel composition of the $\gamma \, \gamma$ subclasses have been supplied \cite{CMSnote,atlasnote}. We have modified our fit to use this information consistently with our rescaling procedure.
This procedure replaces our utilization of estimated $gg$ contamination in the $pp \rightarrow  \gamma \, \gamma  \, jj$ signal events. Also note that
despite the cuts of the ``tight" and ``loose" dijet channels of CMS indicating they are not mutually exclusive event classes, the CMS collaboration vetoes an
event appearing in the ``loose" $pp \rightarrow  \gamma \, \gamma  \, jj$  sample if it passes the ``tight" cuts \cite{privatecomm}. As such, these event classes can both be included in the
global fit we perform.\footnote{We however still thank V. Sanz for kindly providing the contamination coefficients for v1 appropriate for our past procedure.The contamination in the first version of this paper
was reported with a typographical error, the correct contaminations are $\epsilon = 0.032$ for the $7 \, {\rm TeV}$  dijet tagged diphoton signal, $\epsilon = 0.023$ for the $8 \, {\rm TeV}$ ``tight" dijet tagged diphoton data
and $\epsilon = 0.039$ for the $8 \, {\rm TeV}$ ``loose" dijet tagged diphoton data \cite{toappear}. Here $\epsilon$ is the contamination of the $pp \rightarrow  \gamma \, \gamma  \, jj$ signals due to $gg$ Higgs production events,
when $\epsilon$ is defined such that the rate is given by
$\left(\epsilon \, \sigma_{gg\rightarrow h} + \sigma_{jjh} \right) \times {\rm Br}(h \rightarrow \gamma \, \gamma)$.}
Finally we note that the $7,8 \, {\rm TeV}$ signal strengths for the $b\, \bar{b}$ and $ZZ$ signal strengths have now been supplied and are incorporated in our fit.
Comparing our extracted $ZZ$ $8 \, {\rm TeV}$ result to the experimentally supplied number we find agreement within the estimated accuracy of our procedure.
For the $b \, \bar{b}$ CMS signal strength we note that the $7 \, {\rm TeV}$ signal strength recently reported in Ref.~\cite{talkfriday} differs from the previously public
$7 \, {\rm TeV}$ signal strength.

\subsection*{Acknowledgments}
We thank A.~Djouadi, J.~Erler, B.~Feigl, R.~Gon\c{c}alo, R.~Harlander, A.~Juste, M.~Martinez, M.~Schumacher, M.~Spira, W.~Fisher, J.~Huston, V.~Sanz,
V.~Sharma, P.~Uwer, J.~Bendavid, A. Delgado, K. Tackmann and J. Bergstr\"om for helpful communication on related theory and data.
This work has been partly supported by the European Commission under the contract ERC advanced
grant 226371 MassTeV, the contract PITN-GA-2009-237920 UNILHC, and
the contract MRTN-CT-2006-035863 ForcesUniverse, as well as by the
Spanish Consolider Ingenio 2010 Programme CPAN (CSD2007-00042) and the
Spanish Ministry MICNN under contract FPA2010-17747 and
FPA2008-01430. MM is supported by the DFG SFB/TR9 Computational Particle Physics.
Preprints: KA-TP-29-2012, SFB/CPP-12-48, CERN-PH-TH/2012-151. 

\end{document}